\documentclass[aps, prl, twocolumn, superscriptaddress, amsmath, showpacs, tightenlines, footinbib, longbibliography]{revtex4-2}
\usepackage{graphicx,subfigure,amsmath}
\usepackage[bookmarks=false]{hyperref}
\hypersetup{colorlinks=true,citecolor=blue,linkcolor=blue,urlcolor=blue,pdfstartview=FitH,bookmarksopen=true}

\usepackage{amsmath,amstext}
\usepackage{booktabs}
\usepackage{multirow}
\usepackage{dcolumn}
\usepackage{mathrsfs}
\usepackage{color}
\usepackage{multirow}
\usepackage{afterpage}
\usepackage{subfigure}
\usepackage{amssymb}
\usepackage{wasysym}
\usepackage{graphicx}
\usepackage{caption}
\usepackage{ragged2e}
\usepackage{colortbl} 
\usepackage[table]{xcolor}
\allowdisplaybreaks

\begin{document}
    \title{Enhancing Nonreciprocity through Squeezing-Induced Symmetry Breaking}
	\author{B.-B. Liu}\thanks{These authors contributed equally to this work.}
    	\affiliation{Quantum Information Institute, School of Physics and Laboratory of Zhongyuan Light, Zhengzhou University, Zhengzhou 450001, China}
     \author{D.-Y. Wang}\thanks{These authors contributed equally to this work.}
	\affiliation{Quantum Information Institute, School of Physics and Laboratory of Zhongyuan Light, Zhengzhou University, Zhengzhou 450001, China}
    \author{J. Tang}
    \affiliation{Key Laboratory of Low-Dimensional Quantum Structures and Quantum Control of Ministry of Education, Department of Physics and Synergetic Innovation Center for Quantum Effects and Applications, Hunan Normal University, Changsha 410081, China}
        \author{G. Chen}\email{chengang971@163.com}
	\affiliation{Quantum Information Institute, School of Physics and Laboratory of Zhongyuan Light, Zhengzhou University, Zhengzhou 450001, China}
    \author{H. Jing}\email{jinghui73@foxmail.com}
    \affiliation{Key Laboratory of Low-Dimensional Quantum Structures and Quantum Control of Ministry of Education, Department of Physics and Synergetic Innovation Center for Quantum Effects and Applications, Hunan Normal University, Changsha 410081, China}
  \affiliation{College of Science, National University of Defense Technology,
Changsha 410073, China}
          \author{Shi-Lei Su}\email{slsu@zzu.edu.cn}
	\affiliation{Quantum Information Institute, School of Physics and Laboratory of Zhongyuan Light, Zhengzhou University, Zhengzhou 450001, China}
    \affiliation{Institute of Quantum Materials and Physics, Henan Academy of Sciences, Zhengzhou 450046, China}
         \author{F. Nori}
    \affiliation{Quantum Information Physics Theory Research Team, Center for Quantum Computing, RIKEN, Wakoshi, Saitama 351-0198, Japan}
\date{\today}
        
\begin{abstract}
Reservoir engineering enables unidirectional energy and signal flow. We establish squeezing-induced symmetry breaking between two cavities as a guiding principle for exponentially amplifying reservoir-mediated nonreciprocity. 
Rather than a simple scaling of the coupling, this mechanism strategically redistributes the squeezing resources to relax experimental requirements, as single-cavity squeezing alone demands a much larger squeezing strength. Moreover, reservoir squeezing does not alter the system symmetry, but reshapes the noise correlations and thereby changes the system dynamics. The proposed mechanism improves the performance of the quantum battery by several orders of magnitude, including stored energy, charging power, and ergotropy, with the analytical expressions provided. Extending to the optical isolation, we observe a second-order exponential enhancement of the output signal. Our results open a new avenue for nonreciprocal quantum information processing and nonreciprocal quantum device design.
\end{abstract}

\maketitle
\emph{Introduction.---} 
Nonreciprocity enables unidirectional energy flow, thereby suppressing back-propagating noise, playing a crucial role in the development of quantum information science~\cite{Jalas2013,Bi2011}. To overcome the limitations associated with conventional magneto-optical-based nonreciprocity~\cite{10.1063/1.126284,Shoji_2014}, various approaches have been proposed. These include proposals based on nonlinearity~\cite{PhysRevLett.121.123601,PhysRevResearch.6.033020,Soljacic:03,PhysRevApplied.13.044070,Pan_2022}, effective gauge fields~\cite{PhysRevLett.111.203901,PhysRevLett.126.123603}, optomechanical interaction~\cite{Ruesink2016,Shen2016,PhysRevLett.102.213903,Hafezi:12}, among others~\cite{doi:10.1126/sciadv.abe8924,Peng2014,Maayani2018,Tokura2018,Estep2014,Yang2024,Zhang2025,PhysRevApplied.10.047001,Wu:24,202309835,Chen2022}. Beyond these coherent approaches, recent advances have exploited loss to induce interference among multiple channels, thereby enabling unidirectional flow of energy~\cite{Huang2021,Li2024,PhysRevLett.124.070402,PhysRevA.107.023703}. Notably, coupling systems to a common reservoir provides a highly controllable route, where the interplay between coherent and dissipative dynamics gives rise to a nonreciprocal interaction over a relatively broad bandwidth~\cite{PhysRevX.5.021025,PRXQuantum.4.010306,Fang2017,PhysRevLett.123.127202,PhysRevLett.132.120401,PhysRevLett.126.223603}. Dissipation-induced nonreciprocity has attracted significant interest for its applications in isolators, circulators, and quantum batteries~\cite{PhysRevLett.132.210402}.

Quantum squeezing, as a fundamental nonclassical resource in quantum optics, has been widely investigated in theory~\cite{Loudon01061987,MA201189,PhysRevA.73.063819,Andersen_2016} and demonstrated in experiment~\cite{PhysRevLett.55.2409,doi:10.1126/science.aac5138,Esteve2008,PhysRevLett.100.033602,joumtsev2011,Safavi-Naeini2013,PhysRevLett.117.110801,Zhang2021,Mehmet:11,Marti2024}. Although its capability to enhance effective coupling strengths through parametric amplification is well-established~\cite{PhysRevLett.120.093601,QIN20241,PhysRevLett.114.093602}, the systematic control of nonreciprocal effects via squeezing constitutes an emerging research frontier. Recently, some reports utilizing squeezing-induced frequency shifts have successfully realized both classical and quantum nonreciprocal phenomena~\cite{PhysRevLett.128.083604,DelBino:18,Wang:23}, yet a general and systematic ``design rule" for amplifying or optimizing reservoir-engineered nonreciprocity has not been established.

In this Letter, we investigate whether the squeezing can enhance dissipation-induced nonreciprocal coupling~(NRC) and reveal the underlying physical mechanism. Starting from a general framework, we derive the effective NRC and demonstrate that the intuitively expected enhancement from squeezing is not universal. Instead, it is activated by symmetry breaking between the squeezing
parameters of the system, which provides useful guidance for the experimental realization of strong NRC and, to some extent, relaxes the requirement on the squeezing strength. Distinct from previous schemes~\cite{PhysRevLett.128.083604,PhysRevA.102.053720,Guimond2020,Kannan2023}, the nonreciprocity in our approach emerges from controllable dissipative coupling. Finally, we demonstrate that this mechanism can be exploited to \emph{exponentially} improve the performance of quantum batteries and optical isolators. Our results reveal that these internal mechanisms go beyond simple parameter rescaling, fundamentally reshape the interaction landscape and internal energy-flow pathways, enabling new phenomena and applications inaccessible to previous linear interference approaches.

\begin{figure}
  \centering
  \includegraphics[width=8.5cm,height=8.5cm]{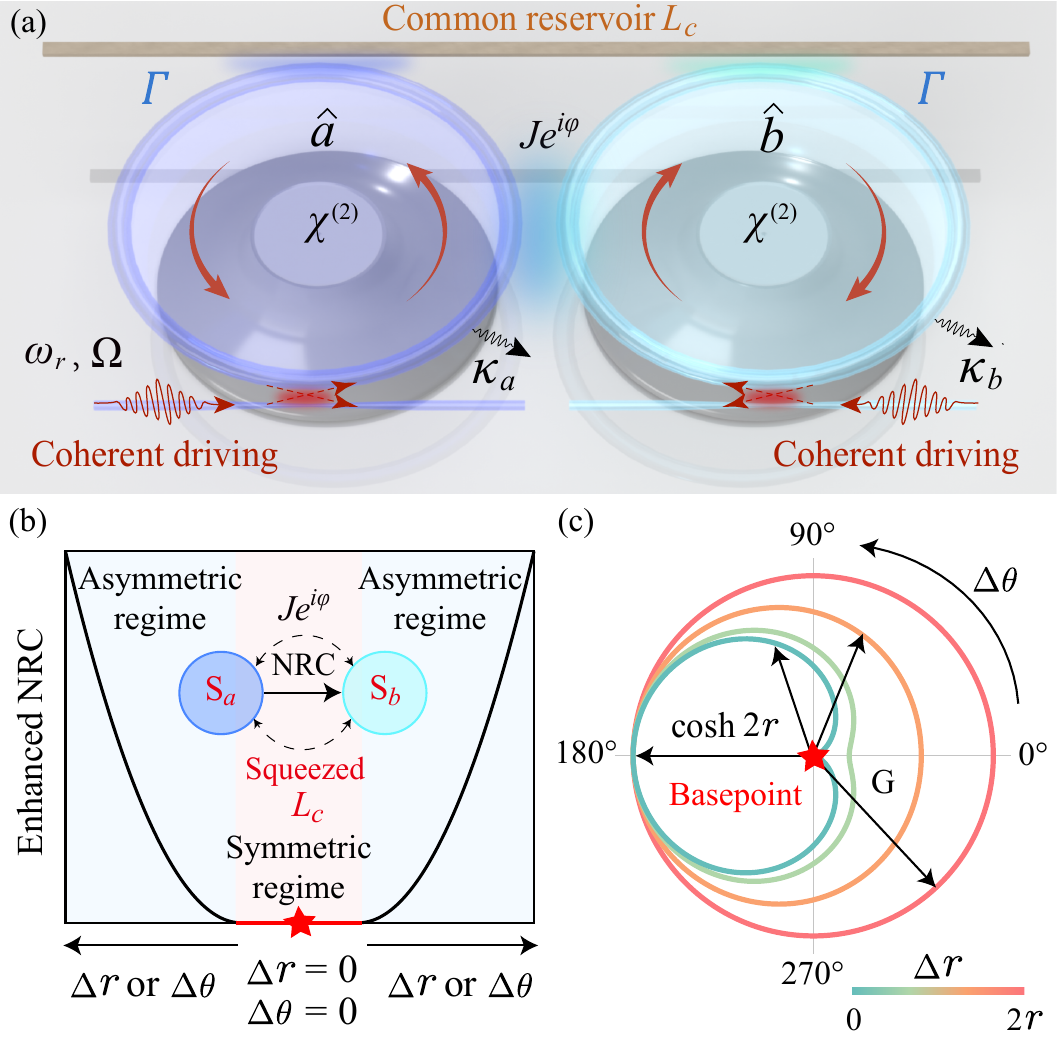} 
  \caption{\justifying (a) Schematic illustration of the nonreciprocal coupling~(NRC) model. Cavity mode $a$ is coherently coupled to cavity mode $b$ with coupling $Je^{i\varphi}$, and both cavity modes are dissipatively coupled into a common reservoir with rate $\Gamma$. (b) Schematic illustration of the effective NRC in exchange-symmetric and asymmetric regimes. $S_{a(b)}$ represents the squeezing applied to cavity mode $a(b)$. (c) Effective NRC enhancement factor $G$ versus the squeezing phase difference $\Delta\theta=\theta_a-\theta_b$ and the amplitude difference $\Delta r=|r_a-r_b|$ between the modes $a$ and $b$, where $r_a+r_b=2r$. The red star marks the non-squeezing case (basepoint).}
  \label{fig1}
\end{figure}

\emph{General theory.---}
We consider a system comprising two cavity modes, $a$ and $b$, with identical resonance frequencies $\omega_a=\omega_b$, coherently coupled with strength $J$ and phase $\varphi$, as illustrated in Fig.~\ref{fig1}(a). In the interaction picture, the Hamiltonian is $H=Je^{i\varphi} a^{\dagger}b+\rm{H.c.}$. Nonreciprocity is engineered by coupling both cavities to a common reservoir. The system dynamics is governed by the master equation $\dot{\rho}(t)=-i[H,\rho(t)]+\sum_{j=a,b,c}\mathcal{L}[L_j]\rho(t)$, where $L_{a(b)}=\sqrt{\kappa_{a(b)}}a(b)$, and $L_c=\sqrt{\Gamma}(p_aa+p_bb)$, with the subscript \(c\) denoting the collective decay channel realized via either the auxiliary cavity or waveguide~\cite{Azouit_2017,7798963,PhysRevA.102.032212,PhysRevA.109.032603,PhysRevA.109.062206,Supplement}. 
The dissipative superoperator $\mathcal{L}[o]\rho=o\rho o^{\dagger}-1/2\{o^{\dagger}o,\rho\}$. Here, $\kappa_{a}$ and $\kappa_{b}$ are the local decay rates, $\Gamma$ is the collective dissipation rate into the reservoir, and the dimensionless coefficients $p_{a, b}$ quantify the relative dissipation of two cavity modes. When the coherent and dissipative interactions are balanced, unidirectional energy transport can be realized~\cite{PhysRevX.5.021025,PhysRevLett.132.210402}. The corresponding effective NRC strength is $2J$, indicated by the red star in Fig.~\ref{fig1}.

We investigate the impact of squeezing on nonreciprocal transport by introducing a $\chi^{(2)}$ nonlinearity. The pump field with resonance frequency $\omega_r$ is subject to the degenerate parametric amplification~(DPA) as it passes through the nonlinear medium. The nonlinear Hamiltonian is $H_{\mathrm{NL}}^h=\Delta h^{\dag}h+\frac{1}{2}\Omega(e^{i\theta_h}h^2+\rm{H.c.})$,
where $\Delta=\omega_h-\frac{1}{2}\omega_r$ is the detuning, $h=a$ or $b$. Here, $\Omega$ and $\theta_h$ denote the coupling strength and phase, respectively. This nonlinear Hamiltonian can be diagonalized via the unitary transformation $S_h = \exp\left[\frac{r_h}{2} e^{-i\theta_h} h^2 - \frac{r_h}{2} e^{i\theta_h} h^{\dag 2} \right]$, yielding $H_{\mathrm{NL}}^h = \omega_s h_s^\dag h_s$, with $h_s=S_h^{\dag}hS_h=\cosh{r_h}h+e^{-i\theta_h}\sinh{r_h}h^{\dag}$. The squeezing parameter is given by $r_h = \frac{1}{4} \ln\left(\frac{1 + \alpha}{1 - \alpha} \right)$, where $\alpha=\Omega/\Delta$, and the squeezed-mode frequency is $\omega_s=\Delta\sqrt{1-\alpha^2}$.

We consider a general scenario where both the cavity modes and the common reservoir are squeezed simultaneously. The Hamiltonian takes the form $H_s=S^{\dag}_bS^{\dag}_a(H+H_{\mathrm{NL}}^a+H_{\mathrm{NL}}^b)S_aS_b$, while the dynamics is governed by the master equation (see the Supplemental Material~\cite{Supplement}, Sec. II) 
\begin{align}
\dot{\rho}_s(t)=&-i[H_s,\rho_s(t)]+\sum_{j=a,b}S_b^{\dag}S_a^{\dag}\mathcal{L}[L_j]\rho(t)S_aS_b\notag\\
&+(N+1)S_b^{\dag}S_a^{\dag}\mathcal{L}[L_c]\rho(t)S_aS_b+NS_b^{\dag}S^{\dag}_a\notag\\
&\times\mathcal{L}[L_c^{\dag}]\rho(t)S_aS_b-MS_b^{\dag}S^{\dag}_a\mathcal{L'}[L_c]\rho(t)S_aS_b\notag\\
&-M^*S_b^{\dag}S^{\dag}_a\mathcal{L'}[L_c^{\dag}]\rho(t)S_aS_b,\label{eq}
\end{align}
where $N=\sinh^2{r_c}$, $M=\cosh{r_c}\sinh{r_c}e^{-i\theta_c}$, and $\mathcal{L'}[o]\rho=o\rho o-1/2\{oo,\rho\}$, with $\{r_c, \theta_c\}$ is the squeezing parameter of the reservoir. To eliminate the additional noise introduced by squeezing, a broadband squeezed-vacuum field is applied to drive the cavity modes $a$ and $b$, such that $S_{a(b)}^{\dag}\mathcal{L}[L_{a(b)}]\rho(t)S_{a(b)}=\mathcal{L}[L_{a_s(b_s)}]\rho_s(t)$~\cite{Ast:13,PhysRevLett.120.093601,PhysRevLett.128.083604}. 
The resulting evolution is
\begin{align}
\frac{d\langle a_s\rangle}{dt}&=-\frac{\Lambda_a+i\omega_s}{2}\langle a_s\rangle+\big(\zeta^*e^{-i\Delta\theta}\sinh{r_a}\sinh{r_b}\notag\\
&~~~~-\zeta\cosh{r_a}\cosh{r_b}\big)\langle b_s\rangle-\big(\zeta^*e^{-i\theta_a}\sinh{r_a}\notag\\
&~~~~\times\cosh{r_b}-\zeta e^{-i\theta_b}\sinh{r_b}\cosh{r_a}\big)\langle b^{\dag}_s\rangle,\notag\\
\frac{d\langle b_s\rangle}{dt}&=-\frac{\Lambda_b+i\omega_s}{2}\langle b_s\rangle+\big(\eta^* e^{i\Delta\theta}\sinh{r_a}\sinh{r_b}\notag\\
&~~~~-\eta\cosh{r_a}\cosh{r_b}\big)\langle a_s\rangle+\big(\eta e^{-i\theta_a}\sinh{r_a}\notag\\
&~~~~\times\cosh{r_b}-\eta^*e^{-i\theta_b}\sinh{r_b}\cosh{r_a}\big)\langle a^{\dag}_s\rangle,\label{eq3}
\end{align}
where $\Lambda_j=\Gamma|p_j|^2+\kappa_j$~($j = a,b$), $\zeta=iJe^{i\varphi}+\mu \frac{\Gamma}{2}$, $\eta=iJe^{-i\varphi}+\mu^* \frac{\Gamma}{2}$, $\Delta\theta=\theta_a-\theta_b$, and $\mu=p_a^{*}p_b$. Nonreciprocity is realized by setting $\varphi=\pm\pi/2$, $\mu=\pm1$, and $J=\frac{\Gamma}{2}$, producing the effective NRC $J_{\rm eff}=\mp2iJ\big(\cosh{r_a}\cosh{r_b}-e^{i\Delta\theta}\sinh{r_a}\sinh{r_b}\big)$. We define the coupling enhancement factor
\begin{align}
G=\left|\frac{J_{\rm eff}}{2J}\right| = \left|\cosh{r_a}\cosh{r_b}-e^{i\Delta\theta}\sinh{r_a}\sinh{r_b}\right|.
\end{align}

Contrary to the intuitive expectation that squeezing always improves system performance, we demonstrate that the amplification of the effective NRC is activated by \emph{breaking the exchange symmetry} between the squeezing parameters of the system, as shown in Fig.~\ref{fig1}(b). When the squeezing parameters are identical~($r_a=r_b,~\theta_a=\theta_b$), the system is symmetric under the exchange of two cavity modes, leading to destructive interference between the squeezing-induced pathways. Consequently, the squeezing contribution cancels, and the effective NRC is reduced to $2J$, identical to the non-squeezing case. Notably, the nonreciprocal phase condition $\varphi=\pm\pi/2$ introduces an overall imaginary phase factor in the effective NRC; therefore, the destructive-interference condition differs from the conventional case $\Delta\theta=\pm\pi$. 

Once this symmetry is broken, the situation changes qualitatively, as illustrated in Fig.~\ref{fig1}(c). For fixed $r_{a(b)}$, the effective NRC increases monotonically with the difference in the squeezing phase difference $\Delta\theta$, reaching its maximum $2J\cosh{2r}$ at $\Delta\theta=\pi$. The behavior for $\pi\leq \Delta\theta\leq2\pi$ mirrors that in $0\leq \Delta\theta\leq\pi$. Similarly, for fixed $\Delta\theta$, the effective NRC increases with the difference in squeezing amplitude $\Delta r$. Notably, the maximal NRC is independent of $\Delta r$ when $\Delta\theta=\pi$, and independent of $\Delta\theta$ when $\Delta r=2r$~($r_a=0$ or $r_b=0$, corresponding to squeezing only applied to one cavity). These results offer practical guidance for experiments, demonstrating that strong NRC can be achieved by controlling the $\Delta\theta$ rather than requiring large squeezing strength. By contrast, when only one cavity is squeezed, achieving the maximal NRC requires $r_a=2r$ or $r_b=2r$, which may be challenging under current experimental constraints.

When only reservoir squeezing is considered, the system satisfies the exchange symmetry. The squeezing parameters $\{r_c,\theta_c\}$ do not appear explicitly in Eq.~(\ref{eq3}), with the operator dynamics becomes
\begin{align}
\frac{d\langle a\rangle}{dt}=-\frac{\Lambda_a}{2}\langle a\rangle-\zeta\langle b\rangle,
~~~\frac{d\langle b\rangle}{dt}=-\frac{\Lambda_b}{2}\langle b\rangle-\eta\langle a\rangle.
\end{align}
Reservoir squeezing reshapes the system–bath interaction by modifying the noise correlation structure. For both vacuum and squeezed reservoirs, the noise has a vanishing mean value, and the influence of reservoir squeezing is solely through second-order correlation functions~\cite{Supplement}. Consequently, it does not modify the effective NRC, remaining identical to the non-squeezing case.

Here, we illustrate the underlying physical mechanisms above. Squeezing cavity modes injects a phase-sensitive gain directly into the coherent coupling path, effectively amplifying the one-way interference that causes nonreciprocal transport. In contrast, squeezing the common reservoir injects energy symmetrically into both modes (as a broadband squeezed vacuum noise). This does not favor one direction over the other at the level of linear dynamics and thus fails to boost the NRC strength. When multiple elements are squeezed, the effective coupling results from a competition among the various squeezing effects. In contrast to the intuitive expectation that any added energy would be beneficial, our analysis shows that only certain pathways benefit from squeezing.

\begin{figure}
	\centering
\includegraphics[width=8.5cm,height=6.5cm]{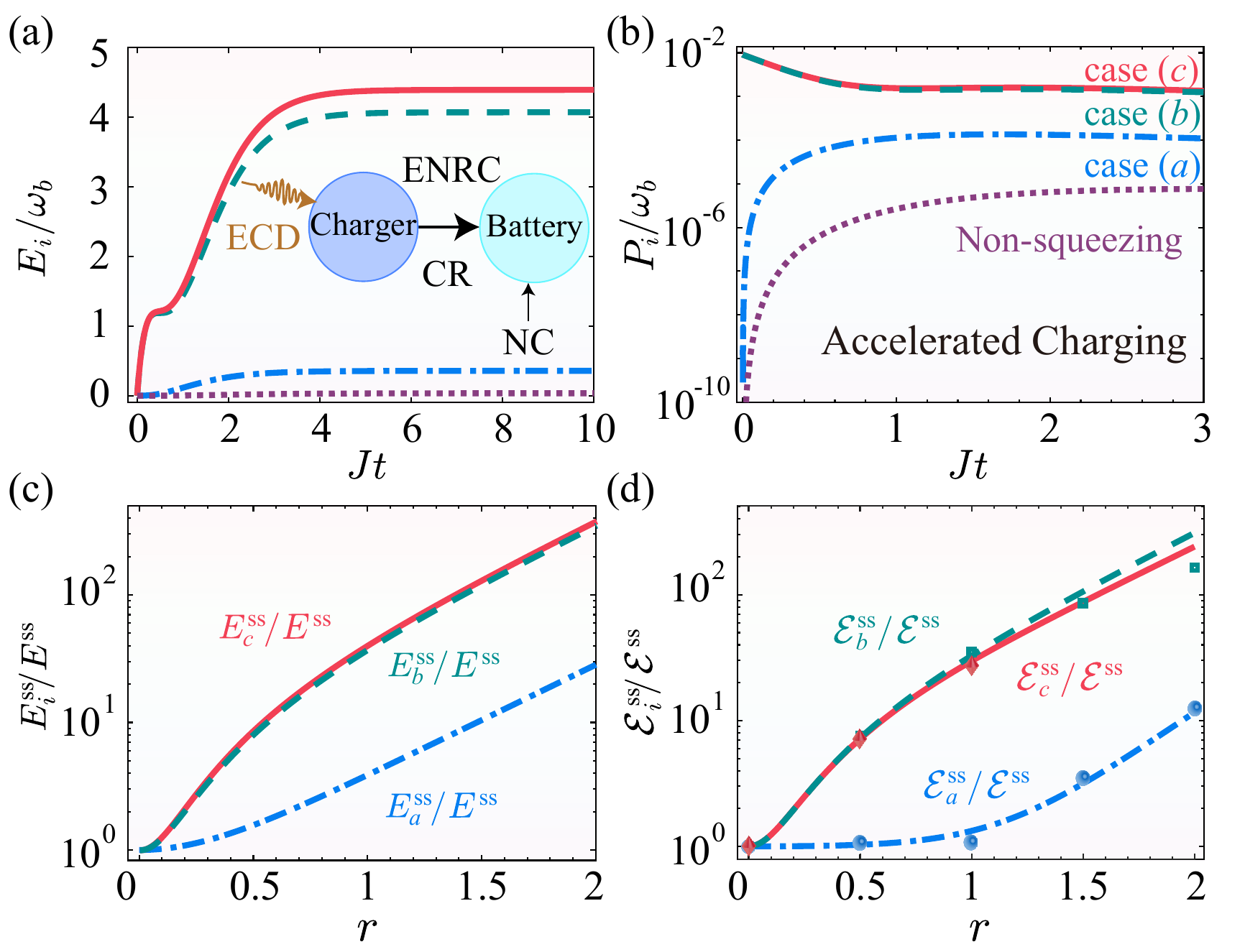}
\caption{\justifying(a) Evolution of the quantum battery energy for four configurations: non-squeezing~\cite{PhysRevLett.132.210402}~(dotted curve), case ($a$)~(dash-dotted curve), case ($b$)~(dashed curve), and case ($c$)~(solid curve), where $r_a=r_c=r=1.5$. Inset: Nonreciprocal charging model and squeezing-induced effects: enhanced classical driving~(ECD), enhanced NRC~(ENRC), counter-rotating term~(CR), and noise correlation~(NC). (b) Average power $P_i$ versus the scaled time $Jt$ during the charging process. (c) Steady-state energy enhancement factor $E_i^{\rm ss}/E^{\rm ss}$ versus the squeezing parameter $r$. (d) Ergotropy enhancement factor $\mathcal{E}_i^{\rm ss}/\mathcal{E}^{\rm ss}$ vs $r$ in the steady state, where $\mathcal{E}^{\rm ss}=E^{\rm ss}$ for the non-squeezing case. The symbols are the numerical results, and the curves are the analytical results. The parameters used are $\kappa=8\times 10^{-5}\omega_b$, $\epsilon= 10^{-4}\omega_b$, $J=10^{-3}\omega_b$, and $\Gamma=2J$.}
	\label{fig2}
\end{figure}

\emph{Quantum battery.---} Quantum batteries have attracted considerable attention~\cite{RevModPhys.96.031001,PhysRevE.87.042123,doi:10.1126/sciadv.abk3160,PhysRevLett.118.150601,PhysRevLett.120.117702,PhysRevLett.122.047702,PhysRevLett.131.240401,PhysRevLett.122.210601,PhysRevLett.132.090401,PhysRevLett.129.130602,d9k1-75d4,PhysRevLett.134.180401,Binder_2015,PhysRevB.105.115405,PhysRevLett.132.210402,73rk-6cp6,67wh-1fxv,jy9l-l8hv}. Reference~\cite{PhysRevLett.132.210402} shows that dissipation-induced nonreciprocity leads to a fourfold enhancement of the stored energy compared to the reciprocal charger–battery model. Our proposed approach can further \emph{exponentially improve} the battery performance, including stored energy, charging power, and ergotropy.
In this context, cavity mode $a$ acts as the charger, while the cavity mode $b$ serves as the quantum battery. Squeezing mode $b$ directly injects energy into the battery, violating the charging principle. Thus, we focus on three cases: squeezing applied to the charger~[case ($a$)], to the common reservoir~[case ($b$)], and simultaneously to the charger and common reservoir~[case ($c$)].

A resonant classical drive field is applied to the charger, described by the Hamiltonian $H_c=\epsilon(a+a^{\dag})$, where $\epsilon$ denotes the driving strength. We assume $\omega_s=0$ and $r_a=r_c$ to simplify the analytical expressions. By solving the evolution equations of the operators, the stored energy $E_i$ ($i=a, b, c$) in the battery for cases ($a$)-($c$) can be obtained~(see the Supplemental Material~\cite{Supplement}, Sec. IV). As shown in Fig.~\ref{fig2}(a), squeezing enables both faster charging and substantially higher energy storage. In case ($a$), squeezing the charger simultaneously enhances the effective NRC, amplifies the classical drive, and induces the counter-rotating term, collectively enabling more efficient energy transfer to the battery. In contrast, reservoir squeezing affects energy transfer by modifying the noise correlations~\cite{Supplement}. In cases ($b$) and ($c$), this leads to an enhanced short-time energy injection with an initial growth rate of $2J\sinh^2{r}$ as $t\rightarrow0$. Meanwhile, squeezing also increases the effective dissipation, causing the system to rapidly reach a transient equilibrium, after which the energy increases until the final steady state is reached. Furthermore, Fig.~\ref{fig2}(b) illustrates that the average charging power is enhanced by \emph{one to two orders of magnitude} compared with the non-squeezing case.

\begin{figure}
	\centering
\includegraphics[width=8.5cm,height=7.13cm]{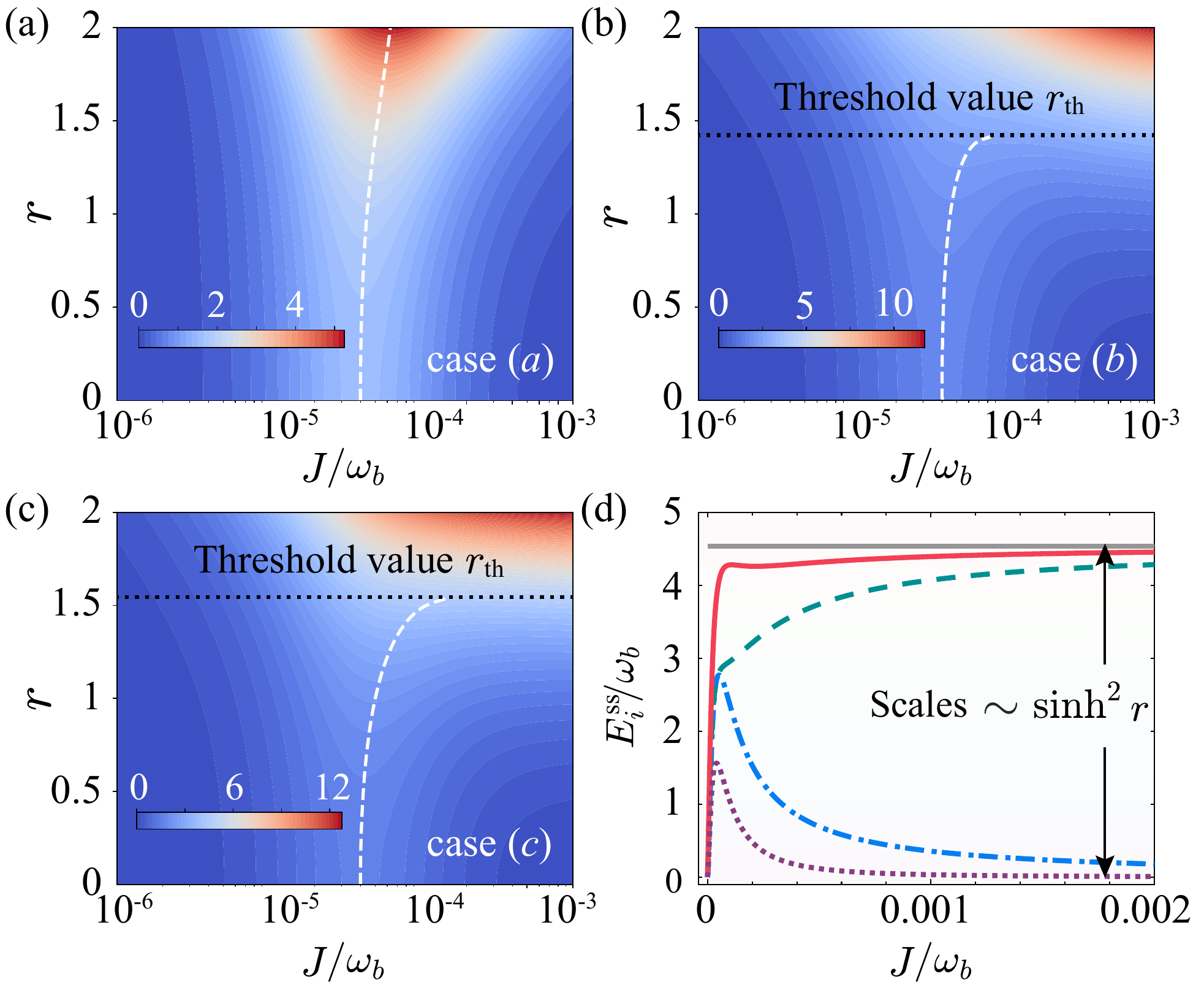}
\caption{\justifying (a)-(c) Steady-state energy of the quantum battery as a function of coupling strength $J$ and squeezing parameter $r$ for case ($a$), case ($b$), and case ($c$). The white dashed line denotes the optimal value $J_{\rm{op}}$ for fixed $r$. (d) Steady-state energy versus $J$ when $r_a=r_c=r=1.5$. The other parameters are $\kappa=8\times 10^{-5}\omega_b$, $\epsilon= 10^{-4}\omega_b$,  $J=10^{-3}\omega_b$, and $\Gamma=2J$.}
	\label{fig22}
\end{figure}

Considering the dissipative dynamics, the system will reach a steady state as $t\rightarrow\infty$. The corresponding energies $E_i^{\rm ss}$ are given by
\begin{align}
E_a^{\rm ss}&=\frac{8J^2\kappa\sinh^2{r}}{(2J+\kappa)^3}+E^{\rm ss}, \notag\\
E_b^{\rm ss}&=\frac{2J(4J^2+\kappa^2)\sinh^2{r}}{(2J+\kappa)^3}+E^{\rm ss},\notag\\
E_c^{\rm ss}&=\frac{2J\sinh^2{r}}{2J+\kappa}+E^{\rm ss},
\end{align}
where we have set $\kappa_a=\kappa_b=\kappa$, and $E^{\rm ss}=64J^2\epsilon^2/(2J+\kappa)^4$ is the steady-state energy for the non-squeezing case. These energies satisfy the relation $E_a^{\rm ss}+E_b^{\rm ss}-E^{\rm ss}=E_c^{\rm ss}$.

The steady-state energy increases quadratically with the classical drive amplitude $\epsilon$, while the squeezing-induced contribution scales as $\sinh^2 r$, as shown in Fig.~\ref{fig2}(c). We emphasize that this monotonic increase is within the ideal broadband Markovian squeezed-reservoir, for which no intrinsic saturation occurs with increasing \(r\). The wide squeezing range simulated here covers most experimentally accessible regimes~\cite {DiGiul,PhysRevLett.134.243603,PhysRevLett.117.110801}. Beyond enhancing NRC, this exponential scaling arises from multiple squeezing-induced effects, underscoring the \emph{constructive role of squeezing in energy accumulation} and persisting in both the strong- and weak-coupling regimes~\cite{Supplement}.

The charging dynamics is mediated by dissipative channels, not all the stored energy is extractable. We quantify the usable energy through the ergotropy, $\mathcal{E}_i=\mathrm{Tr}[\rho_b(t)H_b]-\mathrm{Tr}[\widetilde{\rho}_b(t)H_b]$~\cite{Allahverdyan_2004,PhysRevB.99.035421,Downing2023}, where the subscript ``$b$" denotes the battery subsystem, $\widetilde{\rho}_b$ is the passive state, from which no work can be extracted via unitary cyclic processes. Importantly, dissipation here does not imply thermalization: the squeezed reservoir acts as a nonequilibrium environment that generates nonpassive, extractable energy through squeezing-induced correlations. Our analytical and numerical results, shown in Fig.~\ref{fig2}(d), demonstrate an exponential enhancement of ergotropy with increasing squeezing (see the Supplemental Material~\cite{Supplement}, Sec. V). This reveals a general design principle for quantum batteries: nonequilibrium engineered dissipation can enhance extractable work.
 
 In addition, the effective NRC arises from the interplay between the coherent and dissipative interactions; there exists an optimal coupling strength $J_{\rm{op}}$ that maximizes the stored energy~\cite{Supplement}. As illustrated in Fig.~\ref{fig22}(a), when $J\textless J_{\rm{op}}$, the weak nonreciprocal interaction suppresses the efficiency of energy transfer, greatly accumulating dissipation through the local reservoirs (separate dissipation channels with dissipation rate $\kappa$). In contrast, when $J\textgreater J_{\rm{op}}$, energy is predominantly dissipated into the environment via the nonlocal (common) reservoir due to the limitation of the nonreciprocity condition $\Gamma=2J$.
 
Squeezing of the reservoir modifies the noise correlation and reshapes the dissipative channels. The stored energy comes from two channels: coherent coupling and correlation-modified dissipation, which compete with additional losses to determine the steady-state energy. As shown in Fig.~\ref{fig22}(b), there exists a threshold value $r_{\rm{th}}$, which can be determined analytically~\cite{Supplement}. In the strong squeezing regime $r\textgreater r_{\rm{th}}$, noise correlations dominate the charging dynamics, leading to a monotonic increase of the stored energy with $J$. A similar behavior appears in case ($c$), although squeezing the charger introduces additional enhancement mechanisms that shift the threshold to larger values, as shown in Fig.~\ref{fig22}(c). Moreover, for sufficiently large $J$, the energy transferred to the battery via coherent coupling is greatly dissipated into the nonlocal reservoir, such that the stored energy originates exclusively from the correlation-modified dissipation, which scales as $\sinh^2{r}$, as illustrated in Fig.~\ref{fig22}(d).
 
\emph{Optical isolator.---} Optical isolators are indispensable components in integrated optics~\cite{Bi2011,Hua2016,Tian2021,Yu2023}. Our approach further enables an isolator with \emph{amplified} unidirectional transmission. When both cavity modes are squeezed, the corresponding quantum Langevin equations read
\begin{align}
\frac{da_s}{dt}=&-\frac{\Lambda_a+i\omega_s}{2}a_s+\sqrt{\kappa_a}a^{\mathrm{in}},\notag\\
\frac{db_s}{dt}=&-\frac{\Lambda_b+i\omega_s}{2}b_s+2J\big(e^{i\Delta\theta}\sinh{r_a}\sinh{r_b}\notag\\
&-\cosh{r_a}\cosh{r_b}\big)a_s+2J\big(e^{-i\theta_a}\sinh{r_a}\cosh{r_b}\notag\\
&-e^{-i\theta_b}\sinh{r_b}\cosh{r_a}\big)a_s^{\dag}+\sqrt{\kappa_b}b^{\mathrm{in}},
\end{align}
where $a^{\mathrm{in}}$, $b^{\mathrm{in}}$ are the input quantum fields with zero mean values. The key to amplifying the output signal lies in the term $[2J(e^{-i\theta_a}\sinh{r_a}\cosh{r_b}-e^{-i\theta_b}\sinh{r_b}\cosh{r_a})a_s^{\dag}]$, which arises from counter-rotating contributions induced by coherent and dissipative interactions, an effect intrinsically tied to the squeezing process. This mechanism fundamentally differs from Ref.~\cite{PhysRevX.5.021025}, where the amplification arises from the counter-rotating terms in strong coherent coupling. Moreover, our approach has an exponential enhancement.

\begin{figure}
	\centering
\includegraphics[width=8.5cm,height=3.52cm]{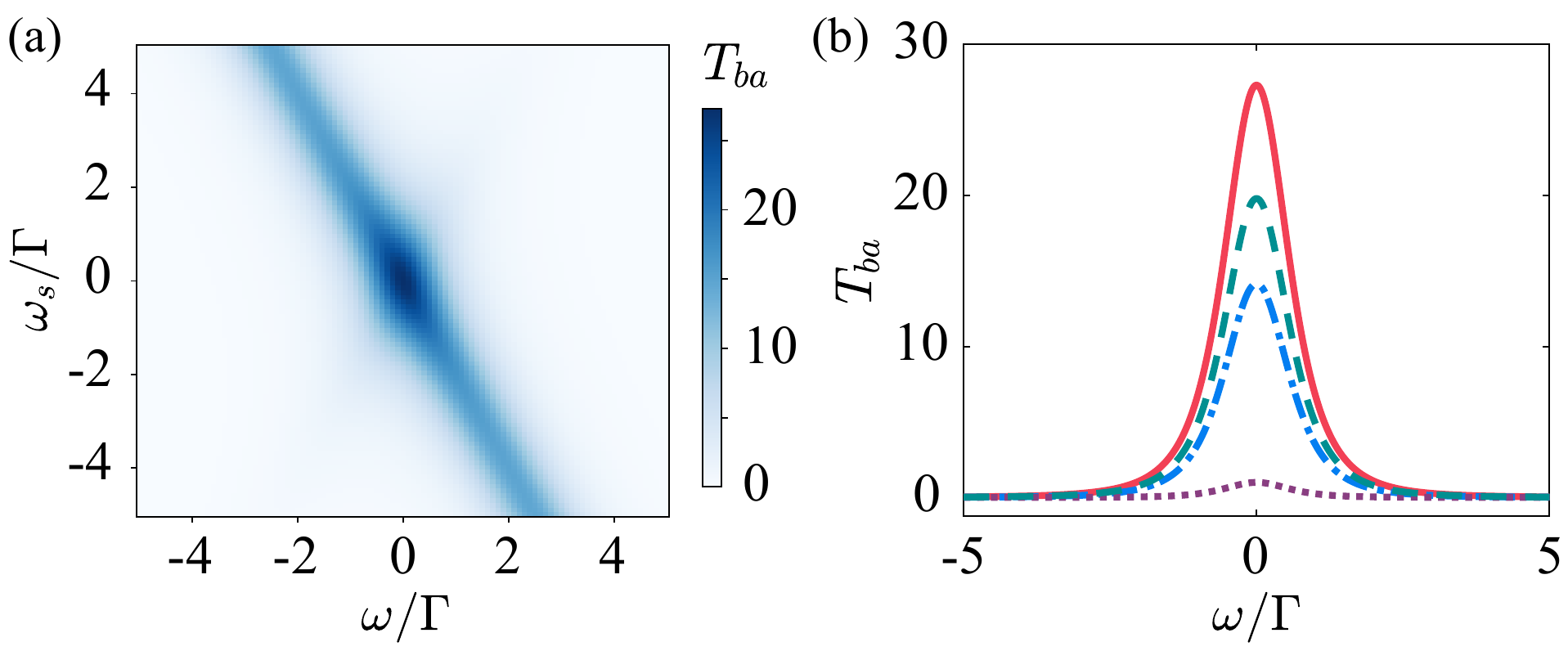}
\caption{\justifying(a) Transmission coefficient $T_{ba}$ versus input signal frequency $\omega$ and squeezed-cavity frequency $\omega_s$, with squeezing parameter $r_a=r_b=1$, and relative phase $\Delta\theta=\pi$. (b) $T_{ba}$ versus $\omega$ for $\omega_s=0$. Here: $r_a=r_b=1$, $\Delta\theta=\pi$ (solid curve); $r_a=0.2$, $r_b=1.8$, $\Delta\theta=\pi/2$ (dashed curve);  $r_a=r_b=1$, $\Delta\theta=\pi/2$ (dash-dotted curve); $r_a=r_b=0$ (dotted curve). Other parameters are $\kappa_a=\kappa_b=8\times10^{-5}\omega_b$, $J=J'_{\mathrm{op}}$, and $\Gamma=2J$.}
    \label{fig3}
\end{figure}

The scattering coefficient matrix can be obtained via Fourier transformation and the input–output relation~\cite{PhysRevA.31.3761,PhysRevA.85.021801}~(see the Supplemental Material~\cite{Supplement}, Sec. VI). The transmission coefficient from mode $b$ to mode $a$ vanishes, $T_{ab}=0$, whereas from mode $a$ to mode $b$ is denoted by $T_{ba}$, which reaches its maximum when the squeezed-cavity frequency is zero, as shown in Fig.~\ref{fig3}(a). When $\omega_s=0$, $T_{ba}$ is given by
\begin{align}
T_{ba}=&\frac{64J^2\kappa_a\kappa_b}{(\Lambda_a^2+4\omega^2)(\Lambda_b^2+4\omega^2)}(\cosh{2r_a}\cosh{2r_b}\notag\\
&-\cos{\Delta\theta\sinh{2r_a}\sinh{2r_b}}).
\end{align}
There exists an optimal coupling strength $J'_{\mathrm{op}}=\sqrt{\kappa_a\kappa_b}/2$ that maximizes the transmission coefficient. Enhanced transmission arises from increasing asymmetry of system squeezing, as illustrated in Fig.~\ref{fig3}(b).

\emph{Proposed experiment.---} We propose to consider a lithium-niobate microring-resonator platform with a large $\chi^{(2)}$ nonlinearity~\cite{Wang:17,Zhu:21,10.1117/1.AP.4.3.034001} and high-$Q$ factors up to $3 \times 10^5$~\cite{Zhang:17,Desiatov:19}. For resonance frequency of $\omega_a/2\pi=\omega_b/2\pi=2.4~\rm{THz}$, the intrinsic loss rate is $\kappa/2\pi=8~\rm{MHz}$. The inter-resonator coupling can be tuned by their separation~\cite{Zhang2018,Li2024}, while the coupling phase can be engineered by thermal contact with a Peltier cell or via a cascade directional coupler with an intermediate propagation path and unidirectional S-bend sections~\cite{Borghi2019,3c01814,10.1063/5.0045228}. The squeezing phase $\theta_h$, crucial in the DPA process, is determined by the pump–cavity relative phase and can be tuned via the pump field phase and stabilized using phase-locking techniques~\cite{s61z-fcyp,PhysRevLett.97.011101,Taguchi}. Tunable dissipative coupling is realized via auxiliary cavity or optical waveguide~\cite{PhysRevA.102.053720,1512289,201100018}, where the broadband squeezing is generated either externally and subsequently injected into the reservoir, or produced directly inside the reservoir through an embedded nonlinear medium~\cite{Takanashi:19,doi:10.1126/science.abo6213,Serkland:95,Ledezma:22}. The distance between the two resonators is chosen as $l=2n\pi/k$ ($n\in\mathbb{Z}$) to preserve the dissipative coupling, while eliminating waveguide-induced coherent interaction~\cite{PhysRevX.5.021025,PhysRevA.102.053720,Supplement}. Alternatively, superconducting coplanar waveguide resonators offer a viable platform~\cite {10.1063/1.4947579,10.1063/5.0124821,10.1063/5.0155213}, with tunable coherent~\cite {5648375,PhysRevResearch.5.033119} and dissipative couplings~\cite{Brown2022,10.1063/1.4953209,PhysRevA.86.012318}. Parametric driving is realized via a superconducting quantum interference device, where a time-dependent flux modulates the resonator frequency to induce nonlinear interactions~\cite{PhysRevLett.120.093602,PhysRevA.101.012348,PhysRevA.85.053825,PhysRevApplied.22.014080}. 

\emph{Conclusion.}---
We investigate the role of squeezing in dissipation-induced NRC, identifying the general conditions under which it enhances nonreciprocity and elucidating the underlying physical mechanisms. Applying this framework to quantum batteries, we show that squeezing enhances both the ergotropy and the charging power, while in optical isolators, it enables exponential amplification of the output signal. Our results illustrate how engineered dissipation and quantum resources (squeezing) can cooperatively control energy flow at the quantum level. This opens prospects for dissipation engineering in a variety of contexts, from protecting quantum information to designing one-way quantum devices.

\begin{acknowledgements} \emph{Acknowledgements.}---
This work was supported by the National Natural Science Foundation of China (Grants No. 12125406, No. 12274376, No. 12575032, No. 12204424, No. U24A2015), and the Natural Science Foundation of Henan Province under Grant (No. 232300421075 and No. 262300421880). H.J. is supported by the Quantum Science and Technology-National Science and Technology Major Project (Grant No. 2024ZD0301000), the National Key R\&D Program (No. 2024YFE0102400), the National Natural Scienc Foundation of China (Grants No. 11935006 and No.12421005), and the Hunan Major Sci-Tech Program (Grant No. 2023ZJ1010). F. N. is supported in part by the Japan Science and Technology Agency (JST) [via the CREST Quantum Frontiers program Grant No. JPMJCR24I2, the Quantum Leap Flagship Program (Q-LEAP), the Moonshot R\&D Grant Number JPMJMS256E, and the ASPIRE program (Grant Number JPMJAP2513)].
\end{acknowledgements}

\emph{Data availability.}--- 
The data that support the findings of this Letter are openly available at~\cite{data}.

\bibliography{REV}

@article{Takanashi:19,
author = {Naoto Takanashi and Wataru Inokuchi and Takahiro Serikawa and Akira Furusawa},
journal = {Opt. Express},
number = {13},
pages = {18900--18909},
publisher = {Optica Publishing Group},
title = {Generation and measurement of a squeezed vacuum up to 100 {MHz} at 1550 nm with a semi-monolithic optical parametric oscillator designed towards direct coupling with waveguide modules},
volume = {27},
month = {Jun},
year = {2019},
url = {https://opg.optica.org/oe/abstract.cfm?URI=oe-27-13-18900},
doi = {10.1364/OE.27.018900},
}

@article{
doi:10.1126/science.abo6213,
author = {Rajveer Nehra  and Ryoto Sekine  and Luis Ledezma  and Qiushi Guo  and Robert M. Gray  and Arkadev Roy  and Alireza Marandi },
title = {Few-cycle vacuum squeezing in nanophotonics},
journal = {Science},
volume = {377},
number = {6612},
pages = {1333-1337},
year = {2022},
doi = {10.1126/science.abo6213},
URL = {https://www.science.org/doi/abs/10.1126/science.abo6213},
}

@article{Taguchi,
author = {Yoshitaka Taguchi and Kenichi Oguchi and Zicong Xu and Donguk Cheon and Shun Takahashi and Yuki Sano and Fumiya Harashima and Yasuyuki Ozeki},
journal = {Opt. Express},
number = {5},
pages = {8002--8014},
publisher = {Optica Publishing Group},
title = {Phase locking of squeezed vacuum generated by a single-pass optical parametric amplifier},
volume = {30},
month = {Feb},
year = {2022},
url = {https://opg.optica.org/oe/abstract.cfm?URI=oe-30-5-8002},
doi = {10.1364/OE.452299},
}

@article{PhysRevLett.97.011101,
  title = {Coherent Control of Vacuum Squeezing in the Gravitational-Wave Detection Band},
  author = {Vahlbruch, Henning and Chelkowski, Simon and Hage, Boris and Franzen, Alexander and Danzmann, Karsten and Schnabel, Roman},
  journal = {Phys. Rev. Lett.},
  volume = {97},
  issue = {1},
  pages = {011101},
  numpages = {4},
  year = {2006},
  month = {Jul},
  publisher = {American Physical Society},
  doi = {10.1103/PhysRevLett.97.011101},
  url = {https://link.aps.org/doi/10.1103/PhysRevLett.97.011101}
}

@article{s61z-fcyp,
  title = {Loss-Tolerant Detection of Squeezed States in the $2\text{ }\text{ }\mathrm{\ensuremath{\mu}}\mathrm{m}$ Region},
  author = {Kwan, K. M. and McRae, T. G. and Qin, J. and Gould, D. W. and Chua, S. S. Y. and Junker, J. and Iden, R. and Adya, V. B. and Yap, M. J. and Slagmolen, B. J. J. and McClelland, D. E. and Ward, R. L.},
  journal = {Phys. Rev. Lett.},
  volume = {136},
  issue = {12},
  pages = {123601},
  numpages = {6},
  year = {2026},
  month = {Mar},
  publisher = {American Physical Society},
  doi = {10.1103/s61z-fcyp},
  url = {https://link.aps.org/doi/10.1103/s61z-fcyp}
}

@article{jy9l-l8hv,
  title = {Efficient charging of multiple open quantum batteries through dissipation and pumping},
  author = {Dias, Josephine and Wang, Hui and Nemoto, Kae and Nori, Franco and Munro, William J.},
  journal = {Phys. Rev. A},
  volume = {113},
  issue = {1},
  pages = {012617},
  numpages = {6},
  year = {2026},
  month = {Jan},
  publisher = {American Physical Society},
  doi = {10.1103/jy9l-l8hv},
  url = {https://link.aps.org/doi/10.1103/jy9l-l8hv}
}

@article{PhysRevA.85.021801,
  title = {Optomechanical systems as single-photon routers},
  author = {Agarwal, G. S. and Huang, Sumei},
  journal = {Phys. Rev. A},
  volume = {85},
  issue = {2},
  pages = {021801},
  numpages = {4},
  year = {2012},
  month = {Feb},
  publisher = {American Physical Society},
  doi = {10.1103/PhysRevA.85.021801},
  url = {https://link.aps.org/doi/10.1103/PhysRevA.85.021801}
}

@article{PhysRevA.31.3761,
  title = {Input and output in damped quantum systems: Quantum stochastic differential equations and the master equation},
  author = {Gardiner, C. W. and Collett, M. J.},
  journal = {Phys. Rev. A},
  volume = {31},
  issue = {6},
  pages = {3761--3774},
  numpages = {0},
  year = {1985},
  month = {Jun},
  publisher = {American Physical Society},
  doi = {10.1103/PhysRevA.31.3761},
  url = {https://link.aps.org/doi/10.1103/PhysRevA.31.3761}
}

@article{PhysRevA.109.062206,
  title = {Complete positivity violation of the reduced dynamics in higher-order quantum adiabatic elimination},
  author = {Tokieda, Masaaki and Elouard, Cyril and Sarlette, Alain and Rouchon, Pierre},
  journal = {Phys. Rev. A},
  volume = {109},
  issue = {6},
  pages = {062206},
  numpages = {25},
  year = {2024},
  month = {Jun},
  publisher = {American Physical Society},
  doi = {10.1103/PhysRevA.109.062206},
  url = {https://link.aps.org/doi/10.1103/PhysRevA.109.062206}
}

@article{PhysRevA.109.032603,
  title = {Adiabatic elimination for composite open quantum systems: Reduced-model formulation and numerical simulations},
  author = {Le R\'egent, Francois-Marie and Rouchon, Pierre},
  journal = {Phys. Rev. A},
  volume = {109},
  issue = {3},
  pages = {032603},
  numpages = {20},
  year = {2024},
  month = {Mar},
  publisher = {American Physical Society},
  doi = {10.1103/PhysRevA.109.032603},
  url = {https://link.aps.org/doi/10.1103/PhysRevA.109.032603}
}

@article{PhysRevA.102.032212,
  title = {Projection-based adiabatic elimination of bipartite open quantum systems},
  author = {Saideh, Ibrahim and Finkelstein-Shapiro, Daniel and Pullerits, Tonu and Keller, Arne},
  journal = {Phys. Rev. A},
  volume = {102},
  issue = {3},
  pages = {032212},
  numpages = {11},
  year = {2020},
  month = {Sep},
  publisher = {American Physical Society},
  doi = {10.1103/PhysRevA.102.032212},
  url = {https://link.aps.org/doi/10.1103/PhysRevA.102.032212}
}

@INPROCEEDINGS{7798963,
  author={Azouit, R. and Sarlette, A. and Rouchon, P.},
  booktitle={IEEE Conf. on Decision and Control}, 
  title={Adiabatic elimination for open quantum systems with effective Lindblad master equations}, 
  year={2016},
  volume={},
  number={},
  pages={4559-4565},
  doi={10.1109/CDC.2016.7798963}}

@article{Azouit_2017,
doi = {10.1088/2058-9565/aa7f3f},
url = {https://doi.org/10.1088/2058-9565/aa7f3f},
year = {2017},
month = {sep},
publisher = {IOP Publishing},
volume = {2},
number = {4},
pages = {044011},
author = {Azouit, R and Chittaro, F and Sarlette, A and Rouchon, P},
title = {Towards generic adiabatic elimination for bipartite open quantum systems},
journal = {Quantum Sci. Technol.},
}

@Article{Guimond2020,
author={Guimond, P.-O.
and Vermersch, B.
and Juan, M. L.
and Sharafiev, A.
and Kirchmair, G.
and Zoller, P.},
title={A unidirectional on-chip photonic interface for superconducting circuits},
journal={npj Quantum Inf.},
year={2020},
month={Mar},
day={27},
volume={6},
number={1},
pages={32},
issn={2056-6387},
doi={10.1038/s41534-020-0261-9},
url={https://doi.org/10.1038/s41534-020-0261-9}
}

@Article{Kannan2023,
author={Kannan, Bharath
and Almanakly, Aziza
and Sung, Youngkyu
and Di Paolo, Agustin
and Rower, David A.
and Braum{\"u}ller, Jochen
and Melville, Alexander
and Niedzielski, Bethany M.
and Karamlou, Amir
and Serniak, Kyle
and Veps{\"a}l{\"a}inen, Antti
and Schwartz, Mollie E.
and Yoder, Jonilyn L.
and Winik, Roni
and Wang, Joel I-Jan
and Orlando, Terry P.
and Gustavsson, Simon
and Grover, Jeffrey A.
and Oliver, William D.},
title={On-demand directional microwave photon emission using waveguide quantum electrodynamics},
journal={Nat. Phys.},
year={2023},
month={Mar},
day={01},
volume={19},
number={3},
pages={394-400},
issn={1745-2481},
doi={10.1038/s41567-022-01869-5},
url={https://doi.org/10.1038/s41567-022-01869-5}
}

@article{PhysRevA.102.053720,
  title = {Programmable directional emitter and receiver of itinerant microwave photons in a waveguide},
  author = {Gheeraert, Nicolas and Kono, Shingo and Nakamura, Yasunobu},
  journal = {Phys. Rev. A},
  volume = {102},
  issue = {5},
  pages = {053720},
  numpages = {14},
  year = {2020},
  month = {Nov},
  publisher = {American Physical Society},
  doi = {10.1103/PhysRevA.102.053720},
  url = {https://link.aps.org/doi/10.1103/PhysRevA.102.053720}
}

@article{67wh-1fxv,
  title = {Collective enhancement in nonreciprocal multimode quantum batteries},
  author = {Khan, Niaz Ali and Zhang, Xingyu and Huang, Chenlong and Liu, Yue and He, Dahai},
  journal = {Phys. Rev. B},
  volume = {112},
  issue = {10},
  pages = {104318},
  numpages = {9},
  year = {2025},
  month = {Sep},
  publisher = {American Physical Society},
  doi = {10.1103/67wh-1fxv},
  url = {https://link.aps.org/doi/10.1103/67wh-1fxv}
}

@article{73rk-6cp6,
  title = {Metastability-induced solid-state quantum batteries for powering microwave quantum electronics},
  author = {Wang, Yuanjin and Wu, Hao and Zhao, Qing},
  journal = {Phys. Rev. A},
  volume = {112},
  issue = {3},
  pages = {L030201},
  numpages = {7},
  year = {2025},
  month = {Sep},
  publisher = {American Physical Society},
  doi = {10.1103/73rk-6cp6},
  url = {https://link.aps.org/doi/10.1103/73rk-6cp6}
}

@article{202309835,
author = {Ju, Ran and Cao, Pei-Chao and Wang, Dong and Qi, Minghong and Xu, Liujun and Yang, Shuihua and Qiu, Cheng-Wei and Chen, Hongsheng and Li, Ying},
title = {Nonreciprocal Heat Circulation Metadevices},
journal = {Adv. Mater.},
volume = {36},
number = {3},
pages = {2309835},
doi = {https://doi.org/10.1002/adma.202309835},
url = {https://advanced.onlinelibrary.wiley.com/doi/abs/10.1002/adma.202309835},
year = {2024}
}

@Article{Yang2024,
author={Yang, Shuihua
and Liu, Mengqi
and Zhao, Changying
and Fan, Shanhui
and Qiu, Cheng-Wei},
title={Nonreciprocal thermal photonics},
journal={Nat. Photonics},
year={2024},
month={May},
day={01},
volume={18},
number={5},
pages={412-424},
issn={1749-4893},
doi={10.1038/s41566-024-01409-y},
url={https://doi.org/10.1038/s41566-024-01409-y}
}

@Article{Chen2022,
author={Chen, Yao-Tong
and Du, Lei
and Guo, Lingzhen
and Wang, Zhihai
and Zhang, Yan
and Li, Yong
and Wu, Jin-Hui},
title={Nonreciprocal and chiral single-photon scattering for giant atoms},
journal={Commun. Phys.},
year={2022},
month={Aug},
day={22},
volume={5},
number={1},
pages={215},
issn={2399-3650},
doi={10.1038/s42005-022-00991-3},
url={https://doi.org/10.1038/s42005-022-00991-3}
}

@article{PhysRevA.85.053825,
  title = {Emission spectrum of the driven nonlinear oscillator},
  author = {Andr\'e, Stephan and Guo, Lingzhen and Peano, Vittorio and Marthaler, Michael and Sch\"on, Gerd},
  journal = {Phys. Rev. A},
  volume = {85},
  issue = {5},
  pages = {053825},
  numpages = {7},
  year = {2012},
  month = {May},
  publisher = {American Physical Society},
  doi = {10.1103/PhysRevA.85.053825},
  url = {https://link.aps.org/doi/10.1103/PhysRevA.85.053825}
}

@Article{data,
author={B.-B. Liu
and D.-Y. Wang
and Jian Tang
and Gang Chen
and H. Jing
and Shi-Lei Su
and F. Nori},
title={Data for ``{C}onditional {E}nhancement of {D}issipation-{I}nduced {N}onreciprocity by {Q}uantum
{S}queezing"},
journal={},
doi={https://doi.org/10.5281/zenodo.17231964}
}

@Article{Hua2016,
author={Hua, Shiyue
and Wen, Jianming
and Jiang, Xiaoshun
and Hua, Qian
and Jiang, Liang
and Xiao, Min},
title={Demonstration of a chip-based optical isolator with parametric amplification},
journal={Nat. Commun.},
year={2016},
month={Nov},
day={25},
volume={7},
number={1},
pages={13657},
issn={2041-1723},
doi={10.1038/ncomms13657},
url={https://doi.org/10.1038/ncomms13657}
}

@Article{Tian2021,
author={Tian, Hao
and Liu, Junqiu
and Siddharth, Anat
and Wang, Rui Ning
and Bl{\'e}sin, Terence
and He, Jijun
and Kippenberg, Tobias J.
and Bhave, Sunil A.},
title={Magnetic-free silicon nitride integrated optical isolator},
journal={Nat. Photonics},
year={2021},
month={Nov},
day={01},
volume={15},
number={11},
pages={828-836},
issn={1749-4893},
doi={10.1038/s41566-021-00882-z},
url={https://doi.org/10.1038/s41566-021-00882-z}
}

@Article{Yu2023,
author={Yu, Mengjie
and Cheng, Rebecca
and Reimer, Christian
and He, Lingyan
and Luke, Kevin
and Puma, Eric
and Shao, Linbo
and Shams-Ansari, Amirhassan
and Ren, Xinyi
and Grant, Hannah R.
and Johansson, Leif
and Zhang, Mian
and Lon{\v{c}}ar, Marko},
title={Integrated electro-optic isolator on thin-film lithium niobate},
journal={Nat. Photonics},
year={2023},
month={Aug},
day={01},
volume={17},
number={8},
pages={666-671},
issn={1749-4893},
doi={10.1038/s41566-023-01227-8},
url={https://doi.org/10.1038/s41566-023-01227-8}
}

@article{PhysRevB.105.115405,
  title = {Extended Dicke quantum battery with interatomic interactions and driving field},
  author = {Dou, Fu-Quan and Lu, You-Qi and Wang, Yuan-Jin and Sun, Jian-An},
  journal = {Phys. Rev. B},
  volume = {105},
  issue = {11},
  pages = {115405},
  numpages = {13},
  year = {2022},
  month = {Mar},
  publisher = {American Physical Society},
  doi = {10.1103/PhysRevB.105.115405},
  url = {https://link.aps.org/doi/10.1103/PhysRevB.105.115405}
}

@article{
doi:10.1126/sciadv.abk3160,
author = {James Q. Quach  and Kirsty E. McGhee  and Lucia Ganzer  and Dominic M. Rouse  and Brendon W. Lovett  and Erik M. Gauger  and Jonathan Keeling  and Giulio Cerullo  and David G. Lidzey  and Tersilla Virgili },
title = {Superabsorption in an organic microcavity: Toward a quantum battery},
journal = {Sci. Adv.},
volume = {8},
number = {2},
pages = {eabk3160},
year = {2022},
doi = {10.1126/sciadv.abk3160},
URL = {https://www.science.org/doi/abs/10.1126/sciadv.abk3160},
}

@article{PhysRevE.87.042123,
  title = {Entanglement boost for extractable work from ensembles of quantum batteries},
  author = {Alicki, Robert and Fannes, Mark},
  journal = {Phys. Rev. E},
  volume = {87},
  issue = {4},
  pages = {042123},
  numpages = {4},
  year = {2013},
  month = {Apr},
  publisher = {American Physical Society},
  doi = {10.1103/PhysRevE.87.042123},
  url = {https://link.aps.org/doi/10.1103/PhysRevE.87.042123}
}

@article{Allahverdyan_2004,
doi = {10.1209/epl/i2004-10101-2},
url = {https://dx.doi.org/10.1209/epl/i2004-10101-2},
year = {2004},
month = {aug},
publisher = {},
volume = {67},
number = {4},
pages = {565},
author = {A. E. Allahverdyan and R. Balian and Th. M. Nieuwenhuizen},
title = {Maximal work extraction from finite quantum systems},
journal = {Europhys. Lett.},
}

@article{
doi:10.1126/sciadv.abe8924,
author = {Ming-Xin Dong  and Ke-Yu Xia  and Wei-Hang Zhang  and Yi-Chen Yu  and Ying-Hao Ye  and En-Ze Li  and Lei Zeng  and Dong-Sheng Ding  and Bao-Sen Shi  and Guang-Can Guo  and Franco Nori },
title = {All-optical reversible single-photon isolation at room temperature},
journal = {Sci. Adv.},
volume = {7},
number = {12},
pages = {eabe8924},
year = {2021},
doi = {10.1126/sciadv.abe8924},
URL = {https://www.science.org/doi/abs/10.1126/sciadv.abe8924},
}

@article{QIN20241,
title = {Quantum amplification and simulation of strong and ultrastrong coupling of light and matter},
journal = {Phys. Rep.},
volume = {1078},
pages = {1-59},
year = {2024},
issn = {0370-1573},
doi = {https://doi.org/10.1016/j.physrep.2024.05.003},
url = {https://www.sciencedirect.com/science/article/pii/S0370157324001571},
author = {Wei Qin and Anton Frisk Kockum and Carlos Sánchez Muñoz and Adam Miranowicz and Franco Nori},
}

@Article{Maayani2018,
author={Maayani, Shai
and Dahan, Raphael
and Kligerman, Yuri
and Moses, Eduard
and Hassan, Absar U.
and Jing, Hui
and Nori, Franco
and Christodoulides, Demetrios N.
and Carmon, Tal},
title={Flying couplers above spinning resonators generate irreversible refraction},
journal={Nature},
year={2018},
month={Jun},
day={01},
volume={558},
number={7711},
pages={569-572},
issn={1476-4687},
doi={10.1038/s41586-018-0245-5},
url={https://doi.org/10.1038/s41586-018-0245-5}
}

@Article{Peng2014,
author={Peng, Bo
and {\"O}zdemir, {\c{S}}ahin Kaya
and Lei, Fuchuan
and Monifi, Faraz
and Gianfreda, Mariagiovanna
and Long, Gui Lu
and Fan, Shanhui
and Nori, Franco
and Bender, Carl M.
and Yang, Lan},
title={Parity--time-symmetric whispering-gallery microcavities},
journal={Nat. Phys.},
year={2014},
month={May},
day={01},
volume={10},
number={5},
pages={394-398},
issn={1745-2481},
doi={10.1038/nphys2927},
url={https://doi.org/10.1038/nphys2927}
}

@article{Wu:24,
author = {Haodong Wu and Jiangshan Tang and Mingyuan Chen and Min Xiao and Yanqing Lu and Keyu Xia and Franco Nori},
journal = {Opt. Express},
number = {7},
pages = {11010--11021},
publisher = {Optica Publishing Group},
title = {Passive magnetic-free broadband optical isolator based on unidirectional self-induced transparency},
volume = {32},
month = {Mar},
year = {2024},
url = {https://opg.optica.org/oe/abstract.cfm?URI=oe-32-7-11010},
doi = {10.1364/OE.507019},
}

@article{Pan_2022,
doi = {10.1088/0256-307X/39/12/124201},
url = {https://dx.doi.org/10.1088/0256-307X/39/12/124201},
year = {2022},
month = {nov},
publisher = {Chinese Physical Society and IOP Publishing Ltd},
volume = {39},
number = {12},
pages = {124201},
author = {Pan, Rui-Kai and Tang, Lei and Xia, Keyu and Nori, Franco},
title = {Dynamic Nonreciprocity with a {K}err Nonlinear Resonator},
journal = {Chinese Phys. Lett.},
}

@Article{Esteve2008,
author={Est{\`e}ve, J.
and Gross, C.
and Weller, A.
and Giovanazzi, S.
and Oberthaler, M. K.},
title={Squeezing and entanglement in a {B}ose--{E}instein condensate},
journal={Nature},
year={2008},
month={Oct},
day={01},
volume={455},
number={7217},
pages={1216-1219},
issn={1476-4687},
doi={10.1038/nature07332},
url={https://doi.org/10.1038/nature07332}
}

@article{
doi:10.1126/science.aac5138,
author = {E. E. Wollman  and C. U. Lei  and A. J. Weinstein  and J. Suh  and A. Kronwald  and F. Marquardt  and A. A. Clerk  and K. C. Schwab },
title = {Quantum squeezing of motion in a mechanical resonator},
journal = {Science},
volume = {349},
number = {6251},
pages = {952-955},
year = {2015},
doi = {10.1126/science.aac5138},
URL = {https://www.science.org/doi/abs/10.1126/science.aac5138},
}

@Article{Marti2024,
author={Zhang, Shicheng
and Hu, Yiqi
and Lin, Gongwei
and Niu, Yueping
and Xia, Keyu
and Gong, Jiangbin
and Gong, Shangqing
and Marti, Stefano
and von L{\"u}pke, Uwe
and Joshi, Om
and Yang, Yu
and Bild, Marius
and Omahen, Andraz
and Chu, Yiwen
and Fadel, Matteo},
title={Quantum squeezing in a nonlinear mechanical oscillator},
journal={Nat. Phys.},
year={2024},
month={Sep},
day={01},
volume={20},
number={9},
pages={1448-1453},
issn={1745-2481},
doi={10.1038/s41567-024-02545-6},
url={https://doi.org/10.1038/s41567-024-02545-6}
}

@article{MA201189,
title = {Quantum spin squeezing},
journal = {Phys. Rep.},
volume = {509},
number = {2},
pages = {89-165},
year = {2011},
issn = {0370-1573},
doi = {https://doi.org/10.1016/j.physrep.2011.08.003},
url = {https://www.sciencedirect.com/science/article/pii/S0370157311002201},
author = {Jian Ma and Xiaoguang Wang and C.P. Sun and Franco Nori},
}

@article{Hafezi:12,
author = {Mohammad Hafezi and Peter Rabl},
journal = {Opt. Express},
number = {7},
pages = {7672--7684},
publisher = {Optica Publishing Group},
title = {Optomechanically induced non-reciprocity in microring resonators},
volume = {20},
month = {Mar},
year = {2012},
url = {https://opg.optica.org/oe/abstract.cfm?URI=oe-20-7-7672},
doi = {10.1364/OE.20.007672},
}

@article{PhysRevApplied.10.047001,
  title = {Electromagnetic Nonreciprocity},
  author = {Caloz, Christophe and Al\`u, Andrea and Tretyakov, Sergei and Sounas, Dimitrios and Achouri, Karim and Deck-L\'eger, Zo\'e-Lise},
  journal = {Phys. Rev. Appl.},
  volume = {10},
  issue = {4},
  pages = {047001},
  numpages = {26},
  year = {2018},
  month = {Oct},
  publisher = {American Physical Society},
  doi = {10.1103/PhysRevApplied.10.047001},
  url = {https://link.aps.org/doi/10.1103/PhysRevApplied.10.047001}
}

@Article{Tokura2018,
author={Tokura, Yoshinori
and Nagaosa, Naoto},
title={Nonreciprocal responses from non-centrosymmetric quantum materials},
journal={Nat. Commun.},
year={2018},
month={Sep},
day={14},
volume={9},
number={1},
pages={3740},
issn={2041-1723},
doi={10.1038/s41467-018-05759-4},
url={https://doi.org/10.1038/s41467-018-05759-4}
}

@article{PhysRevLett.121.123601,
  title = {Nonreciprocity Realized with Quantum Nonlinearity},
  author = {Rosario Hamann, Andr\'es and M\"uller, Clemens and Jerger, Markus and Zanner, Maximilian and Combes, Joshua and Pletyukhov, Mikhail and Weides, Martin and Stace, Thomas M. and Fedorov, Arkady},
  journal = {Phys. Rev. Lett.},
  volume = {121},
  issue = {12},
  pages = {123601},
  numpages = {5},
  year = {2018},
  month = {Sep},
  publisher = {American Physical Society},
  doi = {10.1103/PhysRevLett.121.123601},
  url = {https://link.aps.org/doi/10.1103/PhysRevLett.121.123601}
}

@article{PhysRevLett.120.117702,
  title = {High-Power Collective Charging of a Solid-State Quantum Battery},
  author = {Ferraro, Dario and Campisi, Michele and Andolina, Gian Marcello and Pellegrini, Vittorio and Polini, Marco},
  journal = {Phys. Rev. Lett.},
  volume = {120},
  issue = {11},
  pages = {117702},
  numpages = {6},
  year = {2018},
  month = {Mar},
  publisher = {American Physical Society},
  doi = {10.1103/PhysRevLett.120.117702},
  url = {https://link.aps.org/doi/10.1103/PhysRevLett.120.117702}
}

@article{PhysRevLett.118.150601,
  title = {Enhancing the Charging Power of Quantum Batteries},
  author = {Campaioli, Francesco and Pollock, Felix A. and Binder, Felix C. and C\'eleri, Lucas and Goold, John and Vinjanampathy, Sai and Modi, Kavan},
  journal = {Phys. Rev. Lett.},
  volume = {118},
  issue = {15},
  pages = {150601},
  numpages = {6},
  year = {2017},
  month = {Apr},
  publisher = {American Physical Society},
  doi = {10.1103/PhysRevLett.118.150601},
  url = {https://link.aps.org/doi/10.1103/PhysRevLett.118.150601}
}

@article{Binder_2015,
doi = {10.1088/1367-2630/17/7/075015},
url = {https://dx.doi.org/10.1088/1367-2630/17/7/075015},
year = {2015},
month = {jul},
publisher = {IOP Publishing},
volume = {17},
number = {7},
pages = {075015},
author = {Binder, Felix C and Vinjanampathy, Sai and Modi, Kavan and Goold, John},
title = {Quantacell: powerful charging of quantum batteries},
journal = {New J. Phys.},
}

@article{PhysRevLett.122.047702,
  title = {Extractable Work, the Role of Correlations, and Asymptotic Freedom in Quantum Batteries},
  author = {Andolina, Gian Marcello and Keck, Maximilian and Mari, Andrea and Campisi, Michele and Giovannetti, Vittorio and Polini, Marco},
  journal = {Phys. Rev. Lett.},
  volume = {122},
  issue = {4},
  pages = {047702},
  numpages = {5},
  year = {2019},
  month = {Feb},
  publisher = {American Physical Society},
  doi = {10.1103/PhysRevLett.122.047702},
  url = {https://link.aps.org/doi/10.1103/PhysRevLett.122.047702}
}

@article{PhysRevLett.131.240401,
  title = {Charging Quantum Batteries via Indefinite Causal Order: Theory and Experiment},
  author = {Zhu, Gaoyan and Chen, Yuanbo and Hasegawa, Yoshihiko and Xue, Peng},
  journal = {Phys. Rev. Lett.},
  volume = {131},
  issue = {24},
  pages = {240401},
  numpages = {7},
  year = {2023},
  month = {Dec},
  publisher = {American Physical Society},
  doi = {10.1103/PhysRevLett.131.240401},
  url = {https://link.aps.org/doi/10.1103/PhysRevLett.131.240401}
}

@article{PhysRevLett.122.210601,
  title = {Dissipative Charging of a Quantum Battery},
  author = {Barra, Felipe},
  journal = {Phys. Rev. Lett.},
  volume = {122},
  issue = {21},
  pages = {210601},
  numpages = {6},
  year = {2019},
  month = {May},
  publisher = {American Physical Society},
  doi = {10.1103/PhysRevLett.122.210601},
  url = {https://link.aps.org/doi/10.1103/PhysRevLett.122.210601}
}

@article{RevModPhys.96.031001,
  title = {Colloquium: Quantum batteries},
  author = {Campaioli, Francesco and Gherardini, Stefano and Quach, James Q. and Polini, Marco and Andolina, Gian Marcello},
  journal = {Rev. Mod. Phys.},
  volume = {96},
  issue = {3},
  pages = {031001},
  numpages = {30},
  year = {2024},
  month = {Jul},
  publisher = {American Physical Society},
  doi = {10.1103/RevModPhys.96.031001},
  url = {https://link.aps.org/doi/10.1103/RevModPhys.96.031001}
}

@article{PhysRevLett.126.223603,
  title = {Nonreciprocity and Quantum Correlations of Light Transport in Hot Atoms via Reservoir Engineering},
  author = {Lu, Xingda and Cao, Wanxia and Yi, Wei and Shen, Heng and Xiao, Yanhong},
  journal = {Phys. Rev. Lett.},
  volume = {126},
  issue = {22},
  pages = {223603},
  numpages = {6},
  year = {2021},
  month = {Jun},
  publisher = {American Physical Society},
  doi = {10.1103/PhysRevLett.126.223603},
  url = {https://link.aps.org/doi/10.1103/PhysRevLett.126.223603}
}

@Article{Estep2014,
author={Estep, Nicholas A.
and Sounas, Dimitrios L.
and Soric, Jason
and Al{\`u}, Andrea},
title={Magnetic-free non-reciprocity and isolation based on parametrically modulated coupled-resonator loops},
journal={Nat. Phys.},
year={2014},
month={Dec},
day={01},
volume={10},
number={12},
pages={923-927},
issn={1745-2481},
doi={10.1038/nphys3134},
url={https://doi.org/10.1038/nphys3134}
}

@article{Serkland:95,
author = {D. K. Serkland and M. M. Fejer and R. L. Byer and Y. Yamamoto},
journal = {Opt. Lett.},
number = {15},
pages = {1649--1651},
publisher = {Optica Publishing Group},
title = {Squeezing in a quasi-phase-matched $\mathrm{LiNbO}_3$ waveguide},
volume = {20},
month = {Aug},
year = {1995},
url = {https://opg.optica.org/ol/abstract.cfm?URI=ol-20-15-1649},
doi = {10.1364/OL.20.001649},
}

@article{Ledezma:22,
author = {Luis Ledezma and Ryoto Sekine and Qiushi Guo and Rajveer Nehra and Saman Jahani and Alireza Marandi},
journal = {Optica},
number = {3},
pages = {303--308},
publisher = {Optica Publishing Group},
title = {Intense optical parametric amplification in dispersion-engineered nanophotonic lithium niobate waveguides},
volume = {9},
month = {Mar},
year = {2022},
url = {https://opg.optica.org/optica/abstract.cfm?URI=optica-9-3-303},
doi = {10.1364/OPTICA.442332},
}

@article{Supplement,
journal = {See Supplemental Material [URL] for detailed derivations of the Lindblad operator for the common reservoir, the operator evolution equations in the squeezing framework, analysis of the squeezed-cavity frequency, analytical expressions for energy and ergotropy of the quantum battery, and the scattering coefficient matrix of the optical isolator,which includes Refs.~[58, 61-65, 88, 89, 93, 94]
}
}

@Article{Downing2023,
author={Downing, Charles Andrew
and Ukhtary, Muhammad Shoufie},
title={A quantum battery with quadratic driving},
journal={Commun. Phys.},
year={2023},
month={Nov},
day={04},
volume={6},
number={1},
pages={322},
issn={2399-3650},
doi={10.1038/s42005-023-01439-y},
url={https://doi.org/10.1038/s42005-023-01439-y}
}

@article{PhysRevB.99.035421,
  title = {Charger-mediated energy transfer for quantum batteries: An open-system approach},
  author = {Farina, Donato and Andolina, Gian Marcello and Mari, Andrea and Polini, Marco and Giovannetti, Vittorio},
  journal = {Phys. Rev. B},
  volume = {99},
  issue = {3},
  pages = {035421},
  numpages = {15},
  year = {2019},
  month = {Jan},
  publisher = {American Physical Society},
  doi = {10.1103/PhysRevB.99.035421},
  url = {https://link.aps.org/doi/10.1103/PhysRevB.99.035421}
}

@Article{Brown2022,
author={Brown, T.
and Doucet, E.
and Rist{\`e}, D.
and Ribeill, G.
and Cicak, K.
and Aumentado, J.
and Simmonds, R.
and Govia, L.
and Kamal, A.
and Ranzani, L.},
title={Trade off-free entanglement stabilization in a superconducting qutrit-qubit system},
journal={Nat. Commun.},
year={2022},
month={Jul},
day={09},
volume={13},
number={1},
pages={3994},
issn={2041-1723},
doi={10.1038/s41467-022-31638-0},
url={https://doi.org/10.1038/s41467-022-31638-0}
}

@article{PhysRevA.86.012318,
  title = {Engineering two-mode entangled states between two superconducting resonators by dissipation},
  author = {Li, Peng-Bo and Gao, Shao-Yan and Li, Fu-Li},
  journal = {Phys. Rev. A},
  volume = {86},
  issue = {1},
  pages = {012318},
  numpages = {5},
  year = {2012},
  month = {Jul},
  publisher = {American Physical Society},
  doi = {10.1103/PhysRevA.86.012318},
  url = {https://link.aps.org/doi/10.1103/PhysRevA.86.012318}
}

@article{PhysRevResearch.5.033119,
  title = {Exceptional-point-assisted entanglement, squeezing, and reset in a chain of three superconducting resonators},
  author = {Teixeira, Wallace S. and Vadimov, Vasilii and M\"orstedt, Timm and Kundu, Suman and M\"ott\"onen, Mikko},
  journal = {Phys. Rev. Res.},
  volume = {5},
  issue = {3},
  pages = {033119},
  numpages = {13},
  year = {2023},
  month = {Aug},
  publisher = {American Physical Society},
  doi = {10.1103/PhysRevResearch.5.033119},
  url = {https://link.aps.org/doi/10.1103/PhysRevResearch.5.033119}
}

@article{10.1063/1.4947579,
    author = {Adamyan, A. A. and Kubatkin, S. E. and Danilov, A. V.},
    title = {Tunable superconducting microstrip resonators},
    journal = {Appl. Phys. Lett.},
    volume = {108},
    number = {17},
    pages = {172601},
    year = {2016},
    month = {04},
    issn = {0003-6951},
    doi = {10.1063/1.4947579},
    url = {https://doi.org/10.1063/1.4947579},
}

@article{PhysRevApplied.22.014080,
  title = {Flux-coupled tunable superconducting resonator},
  author = {Li, Juliang and Barry, Pete and Cecil, Tom and Lisovenko, Marharyta and Yefremenko, Volodymyr and Wang, Gensheng and Kruhlov, Serhii and Karapetrov, Goran and Chang, Clarence},
  journal = {Phys. Rev. Appl.},
  volume = {22},
  issue = {1},
  pages = {014080},
  numpages = {11},
  year = {2024},
  month = {Jul},
  publisher = {American Physical Society},
  doi = {10.1103/PhysRevApplied.22.014080},
  url = {https://link.aps.org/doi/10.1103/PhysRevApplied.22.014080}
}

@article{10.1063/5.0124821,
    author = {Shi, Lili and Guo, Tingting and Su, Runfeng and Chi, Tianyuan and Sheng, Yifan and Jiang, Junliang and Cao, Chunhai and Wu, Jingbo and Tu, Xuecou and Sun, Guozhu and Chen, Jian and Wu, Peiheng},
    title = {Tantalum microwave resonators with ultra-high intrinsic quality factors},
    journal = {Appl. Phys. Lett.},
    volume = {121},
    number = {24},
    pages = {242601},
    year = {2022},
    month = {12},
    issn = {0003-6951},
    doi = {10.1063/5.0124821},
    url = {https://doi.org/10.1063/5.0124821},
}

@article{10.1063/5.0155213,
    author = {Krasnok, Alex and Dhakal, Pashupati and Fedorov, Arkady and Frigola, Pedro and Kelly, Michael and Kutsaev, Sergey},
    title = {Superconducting microwave cavities and qubits for quantum information systems},
    journal = {Appl. Phys. Rev.},
    volume = {11},
    number = {1},
    pages = {011302},
    year = {2024},
    month = {01},
    issn = {1931-9401},
    doi = {10.1063/5.0155213},
    url = {https://doi.org/10.1063/5.0155213},
}

@article{201100018,
author = {Morichetti, F. and Ferrari, C. and Canciamilla, A. and Melloni, A.},
title = {The first decade of coupled resonator optical waveguides: bringing slow light to applications},
journal = {Laser Photon. Rev.},
volume = {6},
number = {1},
pages = {74-96},
doi = {https://doi.org/10.1002/lpor.201100018},
url = {https://onlinelibrary.wiley.com/doi/abs/10.1002/lpor.201100018},
year = {2012}
}

@ARTICLE{1512289,
  author={Chremmos, I. and Uzunoglu, N.},
  journal={IEEE Photon. Technol. Lett.}, 
  title={Reflective properties of double-ring resonator system coupled to a waveguide}, 
  year={2005},
  volume={17},
  number={10},
  pages={2110-2112},
  doi={10.1109/LPT.2005.854346}}

@article{PhysRevLett.134.180401,
  title = {Topological Quantum Batteries},
  author = {Lu, Zhi-Guang and Tian, Guoqing and L\"u, Xin-You and Shang, Cheng},
  journal = {Phys. Rev. Lett.},
  volume = {134},
  issue = {18},
  pages = {180401},
  numpages = {8},
  year = {2025},
  month = {May},
  publisher = {American Physical Society},
  doi = {10.1103/PhysRevLett.134.180401},
  url = {https://link.aps.org/doi/10.1103/PhysRevLett.134.180401}
}

@article{d9k1-75d4,
  title = {Self-Discharging Mitigated Quantum Battery},
  author = {Song, Wan-Lu and Wang, Ji-Ling and Zhou, Bin and Yang, Wan-Li and An, Jun-Hong},
  journal = {Phys. Rev. Lett.},
  volume = {135},
  issue = {2},
  pages = {020405},
  numpages = {8},
  year = {2025},
  month = {Jul},
  publisher = {American Physical Society},
  doi = {10.1103/d9k1-75d4},
  url = {https://link.aps.org/doi/10.1103/d9k1-75d4}
}

@article{PhysRevLett.132.090401,
  title = {Remote Charging and Degradation Suppression for the Quantum Battery},
  author = {Song, Wan-Lu and Liu, Hai-Bin and Zhou, Bin and Yang, Wan-Li and An, Jun-Hong},
  journal = {Phys. Rev. Lett.},
  volume = {132},
  issue = {9},
  pages = {090401},
  numpages = {8},
  year = {2024},
  month = {Feb},
  publisher = {American Physical Society},
  doi = {10.1103/PhysRevLett.132.090401},
  url = {https://link.aps.org/doi/10.1103/PhysRevLett.132.090401}
}

@Article{Zhang2018,
author={Zhang, Jing
and Peng, Bo
and {\"O}zdemir, {\c{S}}ahin Kaya
and Pichler, Kevin
and Krimer, Dmitry O.
and Zhao, Guangming
and Nori, Franco
and Liu, Yu-Xi
and Rotter, Stefan
and Yang, Lan},
title={A phonon laser operating at an exceptional point},
journal={Nat. Photonics},
year={2018},
month={Aug},
day={01},
volume={12},
number={8},
pages={479-484},
issn={1749-4893},
doi={10.1038/s41566-018-0213-5},
url={https://doi.org/10.1038/s41566-018-0213-5}
}

@article{Ast:13,
author = {Stefan Ast and Moritz Mehmet and Roman Schnabel},
journal = {Opt. Express},
number = {11},
pages = {13572--13579},
publisher = {Optica Publishing Group},
title = {High-bandwidth squeezed light at 1550 nm from a compact monolithic {PPKTP} cavity},
volume = {21},
month = {Jun},
year = {2013},
url = {https://opg.optica.org/oe/abstract.cfm?URI=oe-21-11-13572},
doi = {10.1364/OE.21.013572},
}

@article{Mehmet:11,
author = {Moritz Mehmet and Stefan Ast and Tobias Eberle and Sebastian Steinlechner and Henning Vahlbruch and Roman Schnabel},
journal = {Opt. Express},
number = {25},
pages = {25763--25772},
publisher = {Optica Publishing Group},
title = {Squeezed light at 1550 nm with a quantum noise reduction of 12.3 dB},
volume = {19},
month = {Dec},
year = {2011},
url = {https://opg.optica.org/oe/abstract.cfm?URI=oe-19-25-25763},
doi = {10.1364/OE.19.025763},
}

@Article{Zhang2025,
author={Zhang, Zimo
and Xu, Zhongxiao
and Huang, Ran
and Lu, Xingda
and Zhang, Fengbo
and Li, Donghao
and {\"O}zdemir, {\c{S}}ahin K.
and Nori, Franco
and Bao, Han
and Xiao, Yanhong
and Chen, Bing
and Jing, Hui
and Shen, Heng},
title={Chirality-induced quantum non-reciprocity},
journal={Nat. Photonics},
year={2025},
month={Aug},
day={01},
volume={19},
number={8},
pages={840-846},
issn={1749-4893},
doi={10.1038/s41566-025-01683-4},
url={https://doi.org/10.1038/s41566-025-01683-4}
}

@article{PhysRevLett.102.213903,
  title = {Optical Nonreciprocity in Optomechanical Structures},
  author = {Manipatruni, Sasikanth and Robinson, Jacob T. and Lipson, Michal},
  journal = {Phys. Rev. Lett.},
  volume = {102},
  issue = {21},
  pages = {213903},
  numpages = {4},
  year = {2009},
  month = {May},
  publisher = {American Physical Society},
  doi = {10.1103/PhysRevLett.102.213903},
  url = {https://link.aps.org/doi/10.1103/PhysRevLett.102.213903}
}

@article{PhysRevApplied.13.044070,
  title = {Nonreciprocity via Nonlinearity and Synthetic Magnetism},
  author = {Xu, Xun-Wei and Li, Yong and Li, Baijun and Jing, Hui and Chen, Ai-Xi},
  journal = {Phys. Rev. Appl.},
  volume = {13},
  issue = {4},
  pages = {044070},
  numpages = {13},
  year = {2020},
  month = {Apr},
  publisher = {American Physical Society},
  doi = {10.1103/PhysRevApplied.13.044070},
  url = {https://link.aps.org/doi/10.1103/PhysRevApplied.13.044070}
}

@article{PhysRevLett.123.127202,
  title = {Nonreciprocity and Unidirectional Invisibility in Cavity Magnonics},
  author = {Wang, Yi-Pu and Rao, J. W. and Yang, Y. and Xu, Peng-Chao and Gui, Y. S. and Yao, B. M. and You, J. Q. and Hu, C.-M.},
  journal = {Phys. Rev. Lett.},
  volume = {123},
  issue = {12},
  pages = {127202},
  numpages = {6},
  year = {2019},
  month = {Sep},
  publisher = {American Physical Society},
  doi = {10.1103/PhysRevLett.123.127202},
  url = {https://link.aps.org/doi/10.1103/PhysRevLett.123.127202}
}

@article{PhysRevLett.124.070402,
  title = {Tunable Nonreciprocal Quantum Transport through a Dissipative {A}haronov-{B}ohm Ring in Ultracold Atoms},
  author = {Gou, Wei and Chen, Tao and Xie, Dizhou and Xiao, Teng and Deng, Tian-Shu and Gadway, Bryce and Yi, Wei and Yan, Bo},
  journal = {Phys. Rev. Lett.},
  volume = {124},
  issue = {7},
  pages = {070402},
  numpages = {6},
  year = {2020},
  month = {Feb},
  publisher = {American Physical Society},
  doi = {10.1103/PhysRevLett.124.070402},
  url = {https://link.aps.org/doi/10.1103/PhysRevLett.124.070402}
}

@Article{Fang2017,
author={Fang, Kejie
and Luo, Jie
and Metelmann, Anja
and Matheny, Matthew H.
and Marquardt, Florian
and Clerk, Aashish A.
and Painter, Oskar},
title={Generalized non-reciprocity in an optomechanical circuit via synthetic magnetism and reservoir engineering},
journal={Nat. Phys.},
year={2017},
month={May},
day={01},
volume={13},
number={5},
pages={465-471},
issn={1745-2481},
doi={10.1038/nphys4009},
url={https://doi.org/10.1038/nphys4009}
}

@Article{Bi2011,
author={Bi, Lei
and Hu, Juejun
and Jiang, Peng
and Kim, Dong Hun
and Dionne, Gerald F.
and Kimerling, Lionel C.
and Ross, C. A.},
title={On-chip optical isolation in monolithically integrated non-reciprocal optical resonators},
journal={Nat. Photonics},
year={2011},
month={Dec},
day={01},
volume={5},
number={12},
pages={758-762},
issn={1749-4893},
doi={10.1038/nphoton.2011.270},
url={https://doi.org/10.1038/nphoton.2011.270}
}

@article{PhysRevA.107.023703,
  title = {Perfect nonreciprocity by loss engineering},
  author = {Huang, Xinyao and Liu, Yong-Chun},
  journal = {Phys. Rev. A},
  volume = {107},
  issue = {2},
  pages = {023703},
  numpages = {7},
  year = {2023},
  month = {Feb},
  publisher = {American Physical Society},
  doi = {10.1103/PhysRevA.107.023703},
  url = {https://link.aps.org/doi/10.1103/PhysRevA.107.023703}
}

@article{PhysRevResearch.6.033020,
  title = {Kerr nonlinearity induced nonreciprocity in dissipatively coupled resonators},
  author = {Miao, Qingtian and Agarwal, G. S.},
  journal = {Phys. Rev. Res.},
  volume = {6},
  issue = {3},
  pages = {033020},
  numpages = {9},
  year = {2024},
  month = {Jul},
  publisher = {American Physical Society},
  doi = {10.1103/PhysRevResearch.6.033020},
  url = {https://link.aps.org/doi/10.1103/PhysRevResearch.6.033020}
}

@article{PhysRevLett.117.110801,
  title = {Detection of 15 $\mathrm{dB}$ Squeezed States of Light and their Application for the Absolute Calibration of Photoelectric Quantum Efficiency},
  author = {Vahlbruch, Henning and Mehmet, Moritz and Danzmann, Karsten and Schnabel, Roman},
  journal = {Phys. Rev. Lett.},
  volume = {117},
  issue = {11},
  pages = {110801},
  numpages = {5},
  year = {2016},
  month = {Sep},
  publisher = {American Physical Society},
  doi = {10.1103/PhysRevLett.117.110801},
  url = {https://link.aps.org/doi/10.1103/PhysRevLett.117.110801}
}

@article{Andersen_2016,
doi = {10.1088/0031-8949/91/5/053001},
url = {https://dx.doi.org/10.1088/0031-8949/91/5/053001},
year = {2016},
month = {apr},
publisher = {IOP Publishing},
volume = {91},
number = {5},
pages = {053001},
author = {Andersen, Ulrik L and Gehring, Tobias and Marquardt, Christoph and Leuchs, Gerd},
title = {30 years of squeezed light generation},
journal = {Phys. Scr.},
}

@Article{Zhang2021,
author={Zhang, Y.
and Menotti, M.
and Tan, K.
and Vaidya, V. D.
and Mahler, D. H.
and Helt, L. G.
and Zatti, L.
and Liscidini, M.
and Morrison, B.
and Vernon, Z.},
title={Squeezed light from a nanophotonic molecule},
journal={Nat. Commun.},
year={2021},
month={Apr},
day={14},
volume={12},
number={1},
pages={2233},
issn={2041-1723},
doi={10.1038/s41467-021-22540-2},
url={https://doi.org/10.1038/s41467-021-22540-2}
}

@article{PhysRevA.73.063819,
  title = {Pulsed squeezed light: Simultaneous squeezing of multiple modes},
  author = {Wasilewski, Wojciech and Lvovsky, A. I. and Banaszek, Konrad and Radzewicz, Czes\l{}aw},
  journal = {Phys. Rev. A},
  volume = {73},
  issue = {6},
  pages = {063819},
  numpages = {12},
  year = {2006},
  month = {Jun},
  publisher = {American Physical Society},
  doi = {10.1103/PhysRevA.73.063819},
  url = {https://link.aps.org/doi/10.1103/PhysRevA.73.063819}
}

@article{PhysRevLett.55.2409,
  title = {Observation of Squeezed States Generated by Four-Wave Mixing in an Optical Cavity},
  author = {Slusher, R. E. and Hollberg, L. W. and Yurke, B. and Mertz, J. C. and Valley, J. F.},
  journal = {Phys. Rev. Lett.},
  volume = {55},
  issue = {22},
  pages = {2409--2412},
  numpages = {0},
  year = {1985},
  month = {Nov},
  publisher = {American Physical Society},
  doi = {10.1103/PhysRevLett.55.2409},
  url = {https://link.aps.org/doi/10.1103/PhysRevLett.55.2409}
}

@Article{Safavi-Naeini2013,
author={Safavi-Naeini, Amir H.
and Gr{\"o}blacher, Simon
and Hill, Jeff T.
and Chan, Jasper
and Aspelmeyer, Markus
and Painter, Oskar},
title={Squeezed light from a silicon micromechanical resonator},
journal={Nature},
year={2013},
month={Aug},
day={01},
volume={500},
number={7461},
pages={185-189},
issn={1476-4687},
doi={10.1038/nature12307},
url={https://doi.org/10.1038/nature12307}
}

@article{Loudon01061987,
author = {R. Loudon and P. L. Knight},
title = {Squeezed Light},
journal = {J. Mod. Opt.},
volume = {34},
number = {6-7},
pages = {709--759},
year = {1987},
publisher = {Taylor \& Francis},
doi = {10.1080/09500348714550721},
URL = {      https://doi.org/10.1080/09500348714550721}
}

@Article{joumtsev2011,
author={Ourjoumtsev, A.
and Kubanek, A.
and Koch, M.
and Sames, C.
and Pinkse, P. W. H.
and Rempe, G.
and Murr, K.},
title={Observation of squeezed light from one atom excited with two photons},
journal={Nature},
year={2011},
month={Jun},
day={01},
volume={474},
number={7353},
pages={623-626},
issn={1476-4687},
doi={10.1038/nature10170},
url={https://doi.org/10.1038/nature10170}
}

@article{PhysRevLett.100.033602,
  title = {Observation of Squeezed Light with 10-$\mathrm{dB}$ Quantum-Noise Reduction},
  author = {Vahlbruch, Henning and Mehmet, Moritz and Chelkowski, Simon and Hage, Boris and Franzen, Alexander and Lastzka, Nico and Go\ss{}ler, Stefan and Danzmann, Karsten and Schnabel, Roman},
  journal = {Phys. Rev. Lett.},
  volume = {100},
  issue = {3},
  pages = {033602},
  numpages = {4},
  year = {2008},
  month = {Jan},
  publisher = {American Physical Society},
  doi = {10.1103/PhysRevLett.100.033602},
  url = {https://link.aps.org/doi/10.1103/PhysRevLett.100.033602}
}

@article{PhysRevA.101.012348,
  title = {Dissipative generation of steady-state squeezing of superconducting resonators via parametric driving},
  author = {Xie, Ji-Kun and Ma, Sheng-Ki and Ren, Ya-Long and Li, Xin-Ke and Li, Fu-Li},
  journal = {Phys. Rev. A},
  volume = {101},
  issue = {1},
  pages = {012348},
  numpages = {7},
  year = {2020},
  month = {Jan},
  publisher = {American Physical Society},
  doi = {10.1103/PhysRevA.101.012348},
  url = {https://link.aps.org/doi/10.1103/PhysRevA.101.012348}
}

@article{PhysRevLett.120.093602,
  title = {Enhancing Cavity Quantum Electrodynamics via Antisqueezing: Synthetic Ultrastrong Coupling},
  author = {Leroux, C. and Govia, L. C. G. and Clerk, A. A.},
  journal = {Phys. Rev. Lett.},
  volume = {120},
  issue = {9},
  pages = {093602},
  numpages = {6},
  year = {2018},
  month = {Mar},
  publisher = {American Physical Society},
  doi = {10.1103/PhysRevLett.120.093602},
  url = {https://link.aps.org/doi/10.1103/PhysRevLett.120.093602}
}

@article{10.1063/1.4953209,
    author = {Bockstiegel, C. and Wang, Y. and Vissers, M. R. and Wei, L. F. and Chaudhuri, S. and Hubmayr, J. and Gao, J.},
    title = {A tunable coupler for superconducting microwave resonators using a nonlinear kinetic inductance transmission line},
    journal = {Appl. Phys. Lett.},
    volume = {108},
    number = {22},
    pages = {222604},
    year = {2016},
    month = {06},
    issn = {0003-6951},
    doi = {10.1063/1.4953209},
    url = {https://doi.org/10.1063/1.4953209},

}

@ARTICLE{5648375,
  author={Gladchenko, Sergiy and Khalil, Moe and Lobb, C. J. and Wellstood, F. C. and Osborn, Kevin D.},
  journal={IEEE Trans. Appl. Supercond.}, 
  title={Superposition of Inductive and Capacitive Coupling in Superconducting $\mathrm{LC}$ Resonators}, 
  year={2011},
  volume={21},
  number={3},
  pages={875-878},
  doi={10.1109/TASC.2010.2089774}}

@article{10.1117/1.AP.4.3.034001,
author = {Milad Gholipour Vazimali and Sasan Fathpour},
title = {{Applications of thin-film lithium niobate in nonlinear integrated photonics}},
volume = {4},
journal = {Adv. Photonics},
number = {3},
publisher = {SPIE},
pages = {034001},
year = {2022},
doi = {10.1117/1.AP.4.3.034001},
URL = {https://doi.org/10.1117/1.AP.4.3.034001}
}

@article{Zhu:21,
author = {Di Zhu and Linbo Shao and Mengjie Yu and Rebecca Cheng and Boris Desiatov and C. J. Xin and Yaowen Hu and Jeffrey Holzgrafe and Soumya Ghosh and Amirhassan Shams-Ansari and Eric Puma and Neil Sinclair and Christian Reimer and Mian Zhang and Marko Lon\v{c}ar},
journal = {Adv. Opt. Photon.},
number = {2},
pages = {242--352},
publisher = {Optica Publishing Group},
title = {Integrated photonics on thin-film lithium niobate},
volume = {13},
month = {Jun},
year = {2021},
url = {https://opg.optica.org/aop/abstract.cfm?URI=aop-13-2-242},
doi = {10.1364/AOP.411024},
}

@article{Wang:17,
author = {Cheng Wang and Xiao Xiong and Nicolas Andrade and Vivek Venkataraman and Xi-Feng Ren and Guang-Can Guo and Marko Lon\v{c}ar},
journal = {Opt. Express},
number = {6},
pages = {6963--6973},
publisher = {Optica Publishing Group},
title = {Second harmonic generation in nano-structured thin-film lithium niobate waveguides},
volume = {25},
month = {Mar},
year = {2017},
url = {https://opg.optica.org/oe/abstract.cfm?URI=oe-25-6-6963},
doi = {10.1364/OE.25.006963},
}

@article{Desiatov:19,
author = {Boris Desiatov and Amirhassan Shams-Ansari and Mian Zhang and Cheng Wang and Marko Lon\v{c}ar},
journal = {Optica},
number = {3},
pages = {380--384},
publisher = {Optica Publishing Group},
title = {Ultra-low-loss integrated visible photonics using thin-film lithium niobate},
volume = {6},
month = {Mar},
year = {2019},
url = {https://opg.optica.org/optica/abstract.cfm?URI=optica-6-3-380},
doi = {10.1364/OPTICA.6.000380},
}

@article{Zhang:17,
author = {Mian Zhang and Cheng Wang and Rebecca Cheng and Amirhassan Shams-Ansari and Marko Lon\v{c}ar},
journal = {Optica},
number = {12},
pages = {1536--1537},
publisher = {Optica Publishing Group},
title = {Monolithic ultra-high-{$Q$} lithium niobate microring resonator},
volume = {4},
month = {Dec},
year = {2017},
url = {https://opg.optica.org/optica/abstract.cfm?URI=optica-4-12-1536},
doi = {10.1364/OPTICA.4.001536},
}

@article{DelBino:18,
author = {Leonardo Del Bino and Jonathan M. Silver and Michael T. M. Woodley and Sarah L. Stebbings and Xin Zhao and Pascal Del'Haye},
journal = {Optica},
number = {3},
pages = {279--282},
publisher = {Optica Publishing Group},
title = {Microresonator isolators and circulators based on the intrinsic nonreciprocity of the Kerr effect},
volume = {5},
month = {Mar},
year = {2018},
url = {https://opg.optica.org/optica/abstract.cfm?URI=optica-5-3-279},
doi = {10.1364/OPTICA.5.000279},
}

@article{Wang:23,
author = {Dong-Yang Wang and Lei-Lei Yan and Shi-Lei Su and Cheng-Hua Bai and Hong-Fu Wang and Erjun Liang},
journal = {Opt. Express},
number = {14},
pages = {22343--22357},
publisher = {Optica Publishing Group},
title = {Squeezing-induced nonreciprocal photon blockade in an optomechanical microresonator},
volume = {31},
month = {Jul},
year = {2023},
url = {https://opg.optica.org/oe/abstract.cfm?URI=oe-31-14-22343},
doi = {10.1364/OE.493208},
}

@Article{Li2024,
author={Li, Bai-Jun
and Zuo, Yun-Lan
and Kuang, Le-Man
and Jing, Hui
and Lee, Chao-Hong},
title={Loss-induced quantum nonreciprocity},
journal={npj Quantum Inf.},
year={2024},
month={Aug},
day={12},
volume={10},
number={1},
pages={75},
issn={2056-6387},
doi={10.1038/s41534-024-00870-5},
url={https://doi.org/10.1038/s41534-024-00870-5}
}

@Article{Huang2021,
author={Huang, Xinyao
and Lu, Cuicui
and Liang, Chao
and Tao, Honggeng
and Liu, Yong-Chun},
title={Loss-induced nonreciprocity},
journal={Light Sci. Appl.},
year={2021},
month={Feb},
day={04},
volume={10},
number={1},
pages={30},
issn={2047-7538},
doi={10.1038/s41377-021-00464-2},
url={https://doi.org/10.1038/s41377-021-00464-2}
}

@Article{Ruesink2016,
author={Ruesink, Freek
and Miri, Mohammad-Ali
and Al{\`u}, Andrea
and Verhagen, Ewold},
title={Nonreciprocity and magnetic-free isolation based on optomechanical interactions},
journal={Nat. Commun.},
year={2016},
month={Nov},
day={29},
volume={7},
number={1},
pages={13662},
issn={2041-1723},
doi={10.1038/ncomms13662},
url={https://doi.org/10.1038/ncomms13662}
}

@article{PhysRevLett.126.123603,
  title = {Synthetic Gauge Fields in a Single Optomechanical Resonator},
  author = {Chen, Yuan and Zhang, Yan-Lei and Shen, Zhen and Zou, Chang-Ling and Guo, Guang-Can and Dong, Chun-Hua},
  journal = {Phys. Rev. Lett.},
  volume = {126},
  issue = {12},
  pages = {123603},
  numpages = {7},
  year = {2021},
  month = {Mar},
  publisher = {American Physical Society},
  doi = {10.1103/PhysRevLett.126.123603},
  url = {https://link.aps.org/doi/10.1103/PhysRevLett.126.123603}
}

@article{PhysRevLett.111.203901,
  title = {Controlling the Flow of Light Using the Inhomogeneous Effective Gauge Field that Emerges from Dynamic Modulation},
  author = {Fang, Kejie and Fan, Shanhui},
  journal = {Phys. Rev. Lett.},
  volume = {111},
  issue = {20},
  pages = {203901},
  numpages = {5},
  year = {2013},
  month = {Nov},
  publisher = {American Physical Society},
  doi = {10.1103/PhysRevLett.111.203901},
  url = {https://link.aps.org/doi/10.1103/PhysRevLett.111.203901}
}

@article{Shoji_2014,
doi = {10.1088/1468-6996/15/1/014602},
url = {https://dx.doi.org/10.1088/1468-6996/15/1/014602},
year = {2014},
month = {jan},
publisher = {IOP Publishing},
volume = {15},
number = {1},
pages = {014602},
author = {Shoji, Yuya and Mizumoto, Tetsuya},
title = {Magneto-optical non-reciprocal devices in silicon photonics},
journal = {Sci. Tech. Adv. Mater.},
}

@Article{Shen2016,
author={Shen, Zhen
and Zhang, Yan-Lei
and Chen, Yuan
and Zou, Chang-Ling
and Xiao, Yun-Feng
and Zou, Xu-Bo
and Sun, Fang-Wen
and Guo, Guang-Can
and Dong, Chun-Hua},
title={Experimental realization of optomechanically induced non-reciprocity},
journal={Nat. Photonics},
year={2016},
month={Oct},
day={01},
volume={10},
number={10},
pages={657-661},
issn={1749-4893},
doi={10.1038/nphoton.2016.161},
url={https://doi.org/10.1038/nphoton.2016.161}
}

@article{Soljacic:03,
author = {Marin Solja\v{c}i\'{c} and Chiyan Luo and J. D. Joannopoulos and Shanhui Fan},
journal = {Opt. Lett.},
number = {8},
pages = {637--639},
publisher = {Optica Publishing Group},
title = {Nonlinear photonic crystal microdevices for optical integration},
volume = {28},
month = {Apr},
year = {2003},
url = {https://opg.optica.org/ol/abstract.cfm?URI=ol-28-8-637},
doi = {10.1364/OL.28.000637},
}

@article{10.1063/1.126284,
    author = {Fujita, J. and Levy, M. and Osgood Jr., R. M.  and Wilkens, L. and Dötsch, H.},
    title = {Waveguide optical isolator based on {M}ach–{Z}ehnder interferometer},
    journal = {Appl. Phys. Lett.},
    volume = {76},
    number = {16},
    pages = {2158-2160},
    year = {2000},
    month = {04},
    issn = {0003-6951},
    doi = {10.1063/1.126284},
    url = {https://doi.org/10.1063/1.126284},
}

@Article{Jalas2013,
author={Jalas, Dirk
and Petrov, Alexander
and Eich, Manfred
and Freude, Wolfgang
and Fan, Shanhui
and Yu, Zongfu
and Baets, Roel
and Popovi{\'{c}}, Milo{\v{s}}
and Melloni, Andrea
and Joannopoulos, John D.
and Vanwolleghem, Mathias
and Doerr, Christopher R.
and Renner, Hagen},
title={What is --- and what is not --- an optical isolator},
journal={Nat. Photonics},
year={2013},
month={Aug},
day={01},
volume={7},
number={8},
pages={579-582},
issn={1749-4893},
doi={10.1038/nphoton.2013.185},
url={https://doi.org/10.1038/nphoton.2013.185}
}

@article{PhysRevLett.128.083604,
  title = {Quantum Squeezing Induced Optical Nonreciprocity},
  author = {Tang, Lei and Tang, Jiangshan and Chen, Mingyuan and Nori, Franco and Xiao, Min and Xia, Keyu},
  journal = {Phys. Rev. Lett.},
  volume = {128},
  issue = {8},
  pages = {083604},
  numpages = {7},
  year = {2022},
  month = {Feb},
  publisher = {American Physical Society},
  doi = {10.1103/PhysRevLett.128.083604},
  url = {https://link.aps.org/doi/10.1103/PhysRevLett.128.083604}
}

@article{PhysRevLett.114.093602,
  title = {Squeezed Optomechanics with Phase-Matched Amplification and Dissipation},
  author = {L\"u, Xin-You and Wu, Ying and Johansson, J. R. and Jing, Hui and Zhang, Jing and Nori, Franco},
  journal = {Phys. Rev. Lett.},
  volume = {114},
  issue = {9},
  pages = {093602},
  numpages = {6},
  year = {2015},
  month = {Mar},
  publisher = {American Physical Society},
  doi = {10.1103/PhysRevLett.114.093602},
  url = {https://link.aps.org/doi/10.1103/PhysRevLett.114.093602}
}

@article{PhysRevLett.120.093601,
  title = {Exponentially Enhanced Light-Matter Interaction, Cooperativities, and Steady-State Entanglement Using Parametric Amplification},
  author = {Qin, Wei and Miranowicz, Adam and Li, Peng-Bo and L\"u, Xin-You and You, J. Q. and Nori, Franco},
  journal = {Phys. Rev. Lett.},
  volume = {120},
  issue = {9},
  pages = {093601},
  numpages = {7},
  year = {2018},
  month = {Mar},
  publisher = {American Physical Society},
  doi = {10.1103/PhysRevLett.120.093601},
  url = {https://link.aps.org/doi/10.1103/PhysRevLett.120.093601}
}

@article{PhysRevLett.132.120401,
  title = {Quantum Criticality in Open Quantum Spin Chains with Nonreciprocity},
  author = {Begg, Samuel E. and Hanai, Ryo},
  journal = {Phys. Rev. Lett.},
  volume = {132},
  issue = {12},
  pages = {120401},
  numpages = {7},
  year = {2024},
  month = {Mar},
  publisher = {American Physical Society},
  doi = {10.1103/PhysRevLett.132.120401},
  url = {https://link.aps.org/doi/10.1103/PhysRevLett.132.120401}
}

@article{PRXQuantum.4.010306,
  title = {Quantum Nonreciprocal Interactions via Dissipative Gauge Symmetry},
  author = {Wang, Yu-Xin and Wang, Chen and Clerk, Aashish A.},
  journal = {PRX Quantum},
  volume = {4},
  issue = {1},
  pages = {010306},
  numpages = {28},
  year = {2023},
  month = {Jan},
  publisher = {American Physical Society},
  doi = {10.1103/PRXQuantum.4.010306},
  url = {https://link.aps.org/doi/10.1103/PRXQuantum.4.010306}
}

@article{PhysRevX.5.021025,
  title = {Nonreciprocal Photon Transmission and Amplification via Reservoir Engineering},
  author = {Metelmann, A. and Clerk, A. A.},
  journal = {Phys. Rev. X},
  volume = {5},
  issue = {2},
  pages = {021025},
  numpages = {16},
  year = {2015},
  month = {Jun},
  publisher = {American Physical Society},
  doi = {10.1103/PhysRevX.5.021025},
  url = {https://link.aps.org/doi/10.1103/PhysRevX.5.021025}
}

@article{PhysRevLett.129.130602,
  title = {Entanglement, Coherence, and Extractable Work in Quantum Batteries},
  author = {Shi, Hai-Long and Ding, Shu and Wan, Qing-Kun and Wang, Xiao-Hui and Yang, Wen-Li},
  journal = {Phys. Rev. Lett.},
  volume = {129},
  issue = {13},
  pages = {130602},
  numpages = {6},
  year = {2022},
  month = {Sep},
  publisher = {American Physical Society},
  doi = {10.1103/PhysRevLett.129.130602},
  url = {https://link.aps.org/doi/10.1103/PhysRevLett.129.130602}
}

@article{PhysRevLett.132.210402,
  title = {Nonreciprocal Quantum Batteries},
  author = {Ahmadi, B. and Mazurek, P. and Horodecki, P. and Barzanjeh, S.},
  journal = {Phys. Rev. Lett.},
  volume = {132},
  issue = {21},
  pages = {210402},
  numpages = {7},
  year = {2024},
  month = {May},
  publisher = {American Physical Society},
  doi = {10.1103/PhysRevLett.132.210402},
  url = {https://link.aps.org/doi/10.1103/PhysRevLett.132.210402}
}

@article{10.1063/5.0045228,
    author = {Liu, Yuzhou G. N. and Hemmatyar, Omid and Hassan, Absar U. and Jung, Pawel S. and Choi, Jae-Hyuck and Christodoulides, Demetrios N. and Khajavikhan, Mercedeh},
    title = {Engineering interaction dynamics in active resonant photonic structures},
    journal = {APL Photonics},
    volume = {6},
    number = {5},
    pages = {050804},
    year = {2021},
    month = {05},
    issn = {2378-0967},
    doi = {10.1063/5.0045228},
    url = {https://doi.org/10.1063/5.0045228},
}

@Article{Borghi2019,
author={Borghi, Massimo
and Trenti, Alessandro
and Pavesi, Lorenzo},
title={Four Wave Mixing control in a photonic molecule made by silicon microring resonators},
journal={Sci. Rep.},
year={2019},
month={Jan},
day={23},
volume={9},
number={1},
pages={408},
issn={2045-2322},
doi={10.1038/s41598-018-36694-5},
url={https://doi.org/10.1038/s41598-018-36694-5}
}

@article{3c01814,
author = {Lim, Min G. and Kim, Dong U. and Park, Young J. and Choi, Dong J. and Jeong, Youngjae and Rah, Yoonhyuk and Hong, Myung S. and Yu, Kyoungsik and Jeong, Kwang-Yong and Han, Sangyoon},
title = {Controlling Four-Wave Mixing through Full Tunability of MEMS-Based Photonic Molecules},
journal = {ACS Photonics},
volume = {11},
number = {9},
pages = {3502-3510},
year = {2024},
doi = {10.1021/acsphotonics.3c01814},
URL = {https://doi.org/10.1021/acsphotonics.3c01814
},
}

@article{DiGiul,
url = {https://doi.org/10.1515/nanoph-2022-0481},
title = {Optical-cavity mode squeezing by free electrons},
title = {},
author = {Valerio Di Giulio and F. Javier García de Abajo},
pages = {4659--4670},
volume = {11},
number = {21},
journal = {Nanophotonics},
doi = {doi:10.1515/nanoph-2022-0481},
year = {2022},
lastchecked = {2026-05-11}
}

@article{PhysRevLett.134.243603,
  title = {Squeezing at the Normal-Mode Splitting Frequency of a Nonlinear Coupled Cavity},
  author = {Junker, Jonas and Qin, Jiayi and Adya, Vaishali B. and Kijbunchoo, Nutsinee and Chua, Sheon S. Y. and McRae, Terry G. and Slagmolen, Bram J. J. and McClelland, David E.},
  journal = {Phys. Rev. Lett.},
  volume = {134},
  issue = {24},
  pages = {243603},
  numpages = {6},
  year = {2025},
  month = {Jun},
  publisher = {American Physical Society},
  doi = {10.1103/PhysRevLett.134.243603},
  url = {https://link.aps.org/doi/10.1103/PhysRevLett.134.243603}
}
	
\clearpage
\onecolumngrid

\section{Supplemental Material}

\setcounter{section}{0}
\setcounter{equation}{0}
\setcounter{figure}{0}
\setcounter{table}{0}

\renewcommand{\thesection}{S\arabic{section}}
\renewcommand{\theequation}{S\arabic{equation}}
\renewcommand{\thefigure}{S\arabic{figure}}
\renewcommand{\thetable}{S\arabic{table}}

\section{Derivation of the effective Lindblad operator for the common reservoir}

Two bosonic modes coupled to a common reservoir exhibit the collective dissipative channel governed by the effective Lindblad operator $L_c=\sqrt{\Gamma}(p_aa+p_bb)$. We show that this mechanism can be implemented either via an auxiliary cavity mode that is adiabatically eliminated, or via directly coupling to a common one-dimensional waveguide.
\subsection{Constructing $L_c$ via the auxiliary cavity mode}
We first consider two cavity modes $a$ and $b$ with the same frequency, interacting through a strongly damped auxiliary cavity mode $z$. In this configuration, the auxiliary mode decays rapidly and continually extracts information from the cavity modes $a$ and $b$. Adiabatic elimination of this fast mode generates an effective Lindblad operator $L_c$. The detailed derivation is as follows. The Hamiltonian of the total system is
\begin{align}
	H^{\rm tot}=H+H^z+H^{\mathrm{int}},
\end{align}
with $H^z=\delta z^{\dag}z$, $H^{\mathrm{int}}=(g_aa+g_bb)z^{\dag}+\mathrm{H.c.}=Qz^{\dag}+\mathrm{H.c.}$, where $\delta$ denotes the detuning between the cavity mode and the reservoir. The dynamics obeys the master equation
\begin{align}
	\dot{\rho}_{\rm tot}=-i[H^{\rm tot},\rho_{\rm tot}]+\kappa_z\mathcal{L}[z]\rho_{\rm tot}=\mathcal{L}\rho_{\rm tot}+\mathcal{L}_z\rho_{\rm tot}+\mathcal{V}\rho_{\rm tot},
\end{align}
where $\mathcal{L}\bullet=-i[H,\bullet]$, $\mathcal{L}_z\bullet=-i[H^z,\bullet]+\kappa_z\mathcal{L}[z]\bullet$, and $\mathcal{V}\bullet=-i[H^{\mathrm{int}},\bullet]$, with $\mathcal{L}[o]\rho=o\rho o^{\dag}-1/2\{o^{\dag}o,\rho\}$. Because the auxiliary mode is strongly dissipative, $\kappa_z\gg\{g_a, g_b, \delta, ||H||\}$, the evolution of the subsystem governed by $H$ is relatively slow, whereas the sector described by $H^z$ relaxes rapidly. The auxiliary cavity relaxes to its unique steady state $\rho_z^{\rm ss}=|0\rangle_z\langle0|$, which satisfies $\mathcal{L}_z\rho_z^{\rm ss}=0$. To eliminate the fast subsystem, we employ the standard adiabatic–elimination expansion $\dot{\rho}=\sum_{i\geq0}L_i\rho$, where $\rho=\mathrm{Tr}_z(\rho_{\rm tot})$ represents the density operator of the system composed of cavity modes $a$ and $b$~\cite{Azouit_2017,7798963,PhysRevA.102.032212}. The zeroth-order contribution vanishes because the fast sector remains in $\rho_z^{\rm ss}$, with $L_0\rho=0$~\cite{PhysRevA.109.032603}. Under the first-order approximation, we obtain~\cite{Azouit_2017}
\begin{align}
	L_1\rho=\mathcal{L}\rho+\mathrm{Tr}_z(\mathcal{V}\rho_{\rm tot})=\mathcal{L}\rho-i\mathrm{Tr}_z([H^{\mathrm{int}},\rho \otimes\rho_z^{\rm ss}])=\mathcal{L}\rho.
\end{align}

Furthermore, the dynamics of the slow subsystem can be represented by a linear, time-invariant Kraus map $\mathcal{K}$, such that the full density matrix satisfies $\rho_{\rm tot}=\mathcal{K}(\rho)$. Within the adiabatic-elimination expansion, the second-order contribution to the effective generator is given by~\cite{Azouit_2017}
\begin{align}
	L_2\rho=\mathrm{Tr}_z(\mathcal{V}(\mathcal{K}_1(\rho))),
\end{align}
with $\mathcal{K}_1(\rho)=\overline{\tau}\overline{\mathcal{K}}_z(\mathcal{V}(\rho\otimes\rho_z^{\rm ss}))+\mathbf{G}_{1,s}(\rho)$. Here $\mathbf{G}_{1,s}$ is the gauge choice associated with the completely-positive, trace-preserving representation for the reduced slow dynamics, where we select $\mathbf{G}_{1,s}(\rho)=-\overline{\tau}\mathrm{Tr}_z(\mathcal{V}(\rho\otimes\rho_z^{\rm ss}))$ for simplicity~\cite{Azouit_2017,PhysRevA.109.062206}. With this choice, we have
\begin{align}
	\mathcal{K}_1(\rho)&=\overline{\tau}\overline{\mathcal{K}}_z(\mathcal{V}(\rho\otimes\rho_z^{\rm ss}))-\overline{\tau}\mathrm{Tr}_z(\mathcal{V}(\rho\otimes\rho_z^{\rm ss}))\notag\\
	&=-i\overline{\tau}\overline{\mathcal{K}}_z(Q\rho\otimes|1\rangle_z\langle0|-\rho Q^{\dag}\otimes|0\rangle_z\langle1|)+i\overline{\tau}\mathrm{Tr}_z(Q\rho\otimes|1\rangle_z\langle0|-\rho Q^{\dag}\otimes|0\rangle_z\langle1|)\notag\\
	&=-i\Big(\frac{Q\rho}{i\delta+\kappa_z/2}\otimes|1\rangle_z\langle0|-\frac{\rho Q^{\dag}}{-i\delta+\kappa_z/2}\otimes|0\rangle_z\langle1|\Big).
\end{align}
Then, the second–order contribution
\begin{align}
	L_2\rho&=\mathrm{Tr}_z(\mathcal{V}(\mathcal{K}_1(\rho)))=-i\mathrm{Tr}_z([H^{\mathrm{int}},\mathcal{K}_1(\rho)])\notag\\
	&=-\frac{Q^{\dag}Q\rho}{i\delta+\kappa_z/2}+\frac{Q\rho Q^{\dag}}{-i\delta+\kappa_z/2}+\frac{Q\rho Q^{\dag}}{i\delta+\kappa_z/2}-\frac{\rho Q^{\dag}Q}{-i\delta+\kappa_z/2}\notag\\
	&=2\alpha\big(Q\rho Q^{\dag}-\frac{1}{2}\{Q^{\dag}Q,\rho\}\big)+i[\delta H,\rho],
\end{align}
where $\alpha=\kappa_z(2\delta^2+\kappa_z^2/2)^{-1}$, $\delta H=\beta Q^{\dag}Q$, and $\beta=\delta[\delta^2+(\kappa_z/2)^2]^{-1}$.

Combining the zeroth-, first-, and second-order contributions in the adiabatic–elimination expansion, we obtain the effective master equation governing the reduced slow subsystem,
\begin{align}
	\dot{\rho}=-i[H-\delta H,\rho]+\Gamma_{\mathrm{eff}}\mathcal{L}[Q]\rho,
\end{align}
with $\Gamma_\mathrm{eff}=2\alpha$.

When the auxiliary mode is resonant with the system, one has $\beta=0$, and therefore $\delta H=0$. In this case, the effective dissipative rate reduces to $\Gamma_{\mathrm{eff}}=4/\kappa_z$. Moreover, the operator $Q$ can be written as $Q=g(p_aa+p_bb)$. As long as both coupling coefficients $g_a$ and $g_b$ are nonzero, they can be rescaled so that $|p_a||p_b|=1$, with the overall magnitude absorbed into the coupling constant $g$. Consequently, the dissipative rate used in the main text is $\Gamma=g\sqrt{\Gamma_{\mathrm{eff}}}$, and the effective Lindblad operator is $L_c=\sqrt{\Gamma}(p_aa+p_bb)$.

\subsection{Constructing $L_c$ via a common waveguide}
We demonstrate that directly coupling two cavity modes to a common 1D waveguide also provides a route to construct the collective dissipation operator $L_c$. Considering two cavity modes $a$ and $b$ at positions $R_a=-d/2$ and $R_b=d/2$, the annihilation operators for the left- and right-propagating modes in the waveguide are denoted by $z_{k,L}$ and $z_{k,R}$. The total Hamiltonian is 
\begin{align}
	H^{\rm tot}=H+H^z+H^{\rm int},
\end{align}
where the waveguide Hamiltonian is 
\begin{align}
	H^z=\sum_k \nu_k(z_{k,L}^{\dag}z_{k,L}+ z_{k,R}^{\dag}z_{k,R}),
\end{align}
with $k$ denotes the wave vector of the propagating mode and $\nu_k$ is the corresponding mode frequency. The interaction Hamiltonian is
\begin{align}
	H^{\rm int}=\sum_{h=a,b}\sum_k\frac{g_k^h}{\sqrt{2}}(h^{\dag}z_{k,R}e^{-i\nu_kt}+hz_{k,R}^{\dag}e^{i\nu_kt}+h^{\dag}z_{k,L}e^{i\nu_kt}+hz_{k,L}^{\dag}e^{-i\nu_kt}).
\end{align}

The evolution of the operator $O$ (cavity modes subspace) in the Heisenberg picture is given by
\begin{align}
	\frac{dO}{dt}=i[H,O]+i\sum_{h=a,b}\sqrt{\frac{\Gamma_h}{2}}([h^{\dag},O]z_{r_h,R}+[h,O]z_{r_h,R}^{\dag}+[h^{\dag},O]z_{r_h,L}+[h,O]z_{r_h,L}^{\dag}).
\end{align}
In deriving the dissipative terms, we have used
\begin{align}
	\sum_k|g_k|^2\int e^{-i(\omega-\nu_k)(t-\tau)}d\tau&=\sum_k2\pi|g_k|^2 \delta(\omega-\nu_k)=L/2\pi\int dk2\pi|g_k|^2\delta(\omega-\nu_k)\notag\\
	&=L/2\pi\int dk2\pi|g_k|^2\delta(\omega-\nu_k)=|g_0|^2/v_g, 
\end{align}
which yields the decay rate 
$\Gamma_h = |g_0^h|^2 / v_g$ under the standard normalization $g_k^h = g_0^h/\sqrt{L}$. 
Here, $v_g = d\nu_k/dk$ denotes the group velocity of the transmission mode. Under the short-delay (Markovian) approximation, where retardation effects are neglected, the free-evolution approximation $h(t-\tau)\approx h(t)e^{i\omega_h\tau}
=h(t)e^{ik_hd}$, where $\tau=d/v_g$ is the propagation delay associated with the cavity separation $d$, and $k_h=\omega_h/v_g$. Together with the slowly-varying decay-rate approximation $\Gamma_h(t-\tau)\approx \Gamma_h(t)$, and the assumption of vacuum input fields for the waveguide. The above formula can be further expressed as~\cite{PhysRevA.102.053720}
\begin{align}
	\frac{dO}{dt}=i[H,O]+\sum_{h=a,b}\frac{\Gamma_h}{2}([h^{\dag},O]h-[h,O]h^{\dag})+\sum_{h\neq h'=a,b}\frac{\sqrt{\Gamma_a\Gamma_b}}{2}([h^{\dag},O]h'e^{ik_{h'}d}-h'^{\dag}[h,O]e^{-ik_{h'}d}).
\end{align}

Accordingly, the expectation value of $O$ can be obtained. Since the above relation holds for an arbitrary operator $O$, the corresponding equation of motion for the density matrix can be directly identified, yielding the master equation
\begin{align}
	\frac{d\rho}{dt}=-i[H+H_J,\rho]+\sum_{h,h'=a,b}\Gamma_{hh'}\mathcal{D}(h,h')\rho,
\end{align}
where the waveguide-mediated exchange interaction between the two cavities is $H_J=J'a^{\dag}b+J'^*b^{\dag}a$, with $J'=-i\sqrt{\Gamma_a\Gamma_b}/4(e^{ik_bd}-e^{-ik_ad})$. Here, $\Gamma_{hh}=\Gamma_h$ denotes the individual decay rate, and $\Gamma_{ab}=\Gamma_{ba}^*=\sqrt{\Gamma_a\Gamma_b}(e^{ik_bd}+e^{-ik_ad})/2$ characterizes the cooperative dissipation. The superoperator is defined as $\mathcal{D}(A, B)\rho=B\rho A^{\dag}-1/2\{A^{\dag}B,\rho\}$. When the separation between the two cavities satisfies $d=2n\pi/k$, with $n\in\mathbb{Z}$, the coherent interaction vanishes, $H_J=0$, while the collective dissipative coupling $L_c=\sqrt{\Gamma}(p_aa+p_bb)$ is constructed, 
\begin{align}
	\frac{d\rho}{dt}=-i[H,\rho]+\mathcal{L}[L_c]\rho,
\end{align}
where we have set $k_a=k_b=k$, $\Gamma_a=\Gamma_b$. 

\section{Evolution of the operators in the squeezing framework} 
In this section, we consider a general scenario where the cavity modes $a$ and $b$ are squeezed with parameters $(r_a, \theta_a)$ and $(r_b, \theta_b)$, respectively, and both are coupled to a common squeezed reservoir characterized by $(r_c, \theta_c)$. 

We first present the derivation of the master equation describing two cavity modes collectively coupled to a squeezing reservoir, using the waveguide as a concrete example. Other interactions are omitted for clarity and discussed later. The interaction Hamiltonian of the cavity modes with the reservoir is
\begin{align}
	H_{\rm int}=\sum_k\big[\frac{g_k^a}{\sqrt{2}}(a^{\dag}z_{k,R}e^{i(\omega_a-\nu_k)t}+a^{\dag}z_{k,L}e^{i(\omega_a+\nu_k)t})+\frac{g_k^b}{\sqrt{2}}(b^{\dag}z_{k,R}e^{i(\omega_b-\nu_k)t}+b^{\dag}z_{k,L}e^{i(\omega_b+\nu_k)t})\big]+\rm{H.c.},
\end{align}
within the Born–Markov approximation, the dynamics of the system is governed by
\begin{align}
	\dot{\rho}(t)=-i\mathrm{Tr}_z[H_\mathrm{int}(t),\rho(0)\otimes\rho_z(0)]-\int_0^td\tau\mathrm{Tr}_z[H_\mathrm{int}(t),[H_\mathrm{int}(\tau),\rho(\tau)\otimes\rho_z(0)]],\label{eqh}
\end{align}
where $\rho_z(0)$ denotes the initial state of the reservoir. 

For the squeezed vacuum reservoir, the density operator is given by 
\begin{align}
	\rho_z=|\xi\rangle\langle\xi|=\prod_kS_k(\xi)|0_k\rangle\langle0_k|S_k^{\dag}(\xi),
\end{align}
where the squeeze operator is
\begin{align}
	S_k(\xi)=\exp\big(\frac{1}{2}\xi^*z_{k_0+k}z_{k_0-k}-\frac{1}{2}\xi z_{k_0+k}^{\dag}z_{k_0-k}^{\dag}\big),
\end{align}
with $\xi=r_c\exp(i\theta_c)$. The correlations can then be obtained:
\begin{align}
	\langle z_k\rangle&=\prod_q\langle0_q|S_q^{\dag}z_kS_q|0_q\rangle=0,\notag\\
	\langle z_k^{\dag}\rangle&=\prod_q\langle0_q|S_q^{\dag}z_k^{\dag}S_q|0_q\rangle=0,\notag\\
	\langle z_k^{\dag}z_{k'}\rangle&=\prod_q\langle0_q|S_q^{\dag}z_k^{\dag}S_qS_q^{\dag}z_{k'}S_q|0_q\rangle=N\delta_{kk'},\notag\\
	\langle z_kz_{k'}^{\dag}\rangle&=\prod_q\langle0_q|S_q^{\dag}z_kS_qS_q^{\dag}z_{k'}^{\dag}S_q|0_q\rangle=(N+1)\delta_{kk'},\notag\\ 
	\langle z_kz_{k'}\rangle&=\prod_q\langle0_q|S_q^{\dag}z_kS_qS_q^{\dag}z_{k'}S_q|0_q\rangle=-M^*\delta_{k',2k_0-k},\notag\\ 
	\langle z_k^{\dag}z_{k'}^{\dag}\rangle&=\prod_q\langle0_q|S_q^{\dag}z_k^{\dag}S_qS_q^{\dag}z_{k'}^{\dag}S_q|0_q\rangle=-M\delta_{k',2k_0-k},
\end{align}
where $N=\sinh^2r_c$, $M=e^{-i\theta_c}\sinh r_c\cosh r_c$, and $k_0=\omega_0/v_g$, with $\omega_0$ is the central frequency of the squeezing device. These correlations are satisfied for both the left- and right-propagating modes.

Substituting $H_{\rm int}$ into Eq.~(\ref{eqh}) and using the aforementioned correlation relationship, we derive the master equation in the squeezed framework
\begin{align}
	\dot{\rho}=&\Gamma_aN(a^{\dag}\rho a-1/2\{a a^{\dag},\rho\})+\Gamma_a(N+1)(a\rho a^{\dag}-1/2\{a^{\dag}a,\rho\})-\Gamma_aM(a\rho a-1/2\{a a,\rho\})\notag\\
	&-\Gamma_aM^*(a^{\dag}\rho a^{\dag}-1/2\{a^{\dag}a^{\dag},\rho\})+\Gamma_bN(b^{\dag}\rho b-1/2\{b b^{\dag},\rho\})+\Gamma_b(N+1)(b\rho b^{\dag}-1/2\{b^{\dag}b,\rho\})\notag\\
	&-\Gamma_bM(b\rho b-1/2\{b b,\rho\})-\Gamma_bM^*(b^{\dag}\rho b^{\dag}-1/2\{b^{\dag}b^{\dag},\rho\})+\Gamma_{ab}N(a^{\dag}\rho b-1/2\{b a^{\dag},\rho\}+b^{\dag}\rho a\notag\\
	&-1/2\{a b^{\dag},\rho\})+\Gamma_{ab}(N+1)(a\rho b^{\dag}-1/2\{b^{\dag} a,\rho\}+b\rho a^{\dag}-1/2\{a^{\dag} b,\rho\})-\Gamma_{ab}M(a\rho b-1/2\{b a,\rho\}\notag\\
	&+b\rho a-1/2\{ab,\rho\})-\Gamma_{ab}M^*(a^{\dag}\rho b^{\dag}-1/2\{b^{\dag} a^{\dag},\rho\}+b^{\dag}\rho a^{\dag}-1/2\{a^{\dag}b^{\dag},\rho\}),
\end{align}
with dissipation rate $\Gamma_h=|g_0^h|^2/v_g$, $\Gamma_{ab}=\sqrt{\Gamma_a\Gamma_b}$. Assuming $\Gamma_a=\Gamma_b$, this expression can be written as
\begin{align}
	\dot{\rho}= (N+1)\mathcal{L}[L_c]\rho+N\mathcal{L}[L_c^{\dag}]\rho-M\mathcal{L}'[L_c]\rho-M^*\mathcal{L}'[L_c^{\dag}]\rho. 
\end{align}

Furthermore, the Hamiltonian of the system in the squeezing framework is $$H_s=S_b^{\dag}S_a^{\dag}(H+H_{\rm NL}^a+H_{\rm NL}^b)S_aS_b=\omega_s(a_s^{\dag}a_s+b_s^{\dag}b_s)+[Je^{i\varphi}(\cosh{r_a}a_s^{\dag}-e^{i\theta_a}\sinh{r_a}a_s)(\cosh{r_b}b_s-e^{-i\theta_b}\sinh{r_b}b^{\dag}_s)+\mathrm{H.c.}],$$ including the local dissipation of the cavity modes, the full master equation can be written as
\begin{align}
	\dot{\rho_s}(t)=&-i[H_s,\rho_s(t)]+\sum_{j=a,b}S_b^{\dag}S_a^{\dag}\mathcal{L}[L_j]\rho(t)S_aS_b+(N+1)S_b^{\dag}S_a^{\dag}\mathcal{L}[L_c]\rho(t)S_aS_b+NS_b^{\dag}S_a^{\dag}\mathcal{L}[L_c^{\dag}]\rho(t)S_aS_b\notag\\
	&-MS_b^{\dag}S_a^{\dag}\mathcal{L'}[L_c]\rho(t)S_aS_b-M^*S_b^{\dag}S_a^{\dag}\mathcal{L'}[L_c^{\dag}]\rho(t)S_aS_b,\notag
\end{align}
which corresponding to Eq.~(1) in the main text. The evolution of the operators is given by the following expressions:
\begin{align}
	\frac{d\langle a_{s}\rangle}{dt}&=-\frac{\Lambda_a+i\omega_s}{2}\langle a_{s}\rangle-(\zeta^*e^{-i\theta_ a}\sinh r_a\cosh r_b-\zeta e^{-i\theta_b}\sinh r_b\cosh r_a)\langle b_{s}^{\dag}\rangle + (\zeta^*e^{-i\Delta\theta}\sinh r_a\sinh r_b\notag\\
	&~~~~-\zeta \cosh r_a\cosh r_b)\langle b_{s}\rangle,\notag\\
	\frac{d\langle b_{s}\rangle}{dt}&=-\frac{\Lambda_b+i\omega_s}{2}\langle b_{s}\rangle+ (\eta e^{-i\theta_a}\sinh r_a\cosh r_b-\eta^*e^{-i\theta_b}\sinh r_b\cosh r_a)\langle a_{s}^{\dag}\rangle +(\eta^*e^{i\Delta\theta} \sinh r_a\sinh r_b\notag\\
	&~~~~-\eta\cosh r_a\cosh r_b)\langle a_{s}\rangle,\notag\\
	\frac{d\langle a_{s}^{\dag}a_{s}\rangle}{dt}&=-\Lambda_a\langle a_{s}^{\dag}a_{s}\rangle+(\zeta e^{-i\theta_b}\sinh r_b\cosh r_a-\zeta^*e^{-i\theta_a}\sinh r_a\cosh r_b)\langle 
	a_{s}^{\dag}b_{s}^{\dag}\rangle+(\zeta^*e^{i\theta_b}\sinh r_b\cosh r_a-\zeta e^{i\theta_a}\notag\\
	&~~~~\times\sinh r_a\cosh r_b)\langle a_{s}b_{s}\rangle-(\zeta \cosh r_a\cosh r_b-\zeta^*e^{-i\Delta\theta}\sinh r_a\sinh r_b)\langle a_{s}^{\dag}b_{s}\rangle+(\zeta e^{i\Delta\theta}\sinh r_a\sinh r_b\notag\\
	&~~~~-\zeta^*\cosh r_a\cosh r_b)\langle a_{s}b_{s}^{\dag}\rangle+\Gamma[ p_a^2e^{-i(\theta_c+\theta_a)}+ p_a^{*2}e^{i(\theta_c+\theta_a)}]\sinh r_a\cosh r_a\sinh r_c\cosh r_c+\Gamma |p_a|^2\notag\\
	&~~~~\times\sinh^2r_a\cosh^2r_c+\Gamma |p_a|^2\cosh^2r_a\sinh^2r_c,\notag\\
	\frac{d\langle a_{s}b_{s}\rangle}{dt}&=-\frac{\Lambda_a+\Lambda_b+4i\omega_s}{2}\langle a_{s}b_{s}\rangle+(\eta e^{-i\theta_a}\sinh r_a\cosh r_b-\eta^*e^{-i\theta_b}\sinh r_b\cosh r_a)\langle 
	a_{s}^{\dag}a_{s}\rangle+(-\zeta^*e^{-i\theta_a}\sinh r_a\notag\\
	&~~~~\times\cosh r_b+\zeta e^{-i\theta_b}\sinh r_b\cosh r_a)\langle 
	b_{s}^{\dag}b_{s}\rangle-(\zeta \cosh r_a\cosh r_b-\zeta^*e^{-i\Delta\theta}\sinh r_a\sinh r_b)\langle 
	b_{s}b_{s}\rangle+(\eta^*e^{i\Delta\theta}\notag\\
	&~~~~\times \sinh r_a\sinh r_b-\eta \cosh r_a\cosh r_b)\langle 
	a_{s}a_{s}\rangle+iJe^{-i(\varphi+\theta_a)}\sinh r_a\cosh r_b+iJe^{i(\varphi-\theta_b)}\sinh r_b\cosh r_a\notag\\
	&~~~~+\frac{\Gamma}{2} p_a p_be^{-i(\theta_c+\theta_a+\theta_b)}\sinh r_a\sinh r_b\sinh 2r_c +\frac{\Gamma}{2}\mu^*e^{-i\theta_a}\sinh r_a\cosh r_b\cosh 2r_c+\frac{\Gamma}{2}p_a^* p_b^*e^{i\theta_c}\cosh r_a\notag\\
	&~~~~\times\cosh r_b\sinh 2r_c+\frac{\Gamma}{2}\mu e^{-i\theta_b}\sinh r_b\cosh r_a\cosh 2r_c,\notag\\
	\frac{d\langle a_{s}^{\dag}b_{s}\rangle}{dt}&=-\frac{\Lambda_a+\Lambda_b}{2}\langle a_{s}^{\dag}b_{s}\rangle+(\zeta^*e^{i\theta_b}\sinh r_b\cosh r_a-\zeta e^{i\theta_a}\sinh r_a\cosh r_b
	)\langle b_{s}b_{s}\rangle+(\eta e^{-i\theta_a}\sinh r_a\cosh r_b-\eta^*\notag\\
	&~~~~\times e^{-i\theta_b}\sinh r_b\cosh r_a)\langle a_s^{\dag}a_s^{\dag}\rangle+(\eta^*e^{i\Delta\theta}\sinh r_a\sinh r_b-\eta\cosh r_a\cosh r_b)\langle a_s^{\dag}a_s\rangle+(\zeta e^{i\Delta\theta}\sinh r_a\notag\\
	&~~~~\times\sinh r_b-\zeta^*\cosh r_a\cosh r_b)\langle b_s^{\dag}b_s\rangle +\Gamma p_ap_be^{-i(\theta_b+\theta_c)}\sinh r_b\cosh r_a\sinh r_c\cosh r_c+\Gamma p_a^*p_b^*e^{i(\theta_a+\theta_c)}\notag\\
	&~~~~\times\sinh r_a\cosh r_b\sinh r_c\cosh r_c+\mu\Gamma e^{i\Delta\theta}\sinh r_a\sinh r_b\cosh^2r_c+\mu^*\Gamma \cosh r_a\cosh r_b\sinh^2r_c,\notag\\
	\frac{d\langle a_sa_s\rangle}{dt}&=-(\Lambda_a+2i\omega_s)\langle a_sa_s\rangle-2(\zeta^*e^{-i\theta_a}\sinh r_a\cosh r_b-\zeta e^{-i\theta_b}\sinh r_b\cosh r_a)\langle a_sb_s^{\dag}\rangle-2(\zeta \cosh r_a\cosh r_b\notag\\
	&~~~~
	+\zeta^* e^{-i\Delta\theta}\sinh r_a\sinh r_b)\langle a_sb_s\rangle+\Gamma p_a^2 e^{-i(2\theta_a+\theta_c)}\sinh^2r_a\sinh r_c\cosh r_c+\Gamma p_a^{*2}e^{i\theta_c}\cosh^2r_a\sinh r_c \notag\\
	&~~~~\times\cosh r_c+\Gamma |p_a|^2e^{-i\theta_a}\sinh r_a\cosh r_a \cosh{2r_c},\notag\\
	\frac{d\langle b_sb_s\rangle}{dt}&=-(\Lambda_b+2i\omega_s)\langle b_sb_s\rangle+2(\eta e^{-i\theta_a}\sinh r_a\cosh r_b-\eta^*e^{-i\theta_b}\sinh r_b\cosh r_a)\langle a_s^{\dag}b_s\rangle-2(\eta \cosh r_a\cosh r_b
	\notag\\
	&~~~~+\eta^*e^{i\Delta\theta}\sinh r_a\sinh r_b)\langle a_sb_s\rangle+\Gamma p_b^2 e^{-i(2\theta_b+\theta_c)}\sinh^2 r_b\sinh r_c\cosh r_c+\Gamma p_b^{*2}e^{i\theta_c}\cosh^2r_b\sinh r_c\notag\\
	&~~~~\times\cosh r_c+\Gamma |p_b|^2e^{-i\theta_b}\sinh r_b\cosh r_b\cosh{2r_c},\notag\\
	\frac{d\langle b_s^{\dag}b_s\rangle}{dt}&=-\Lambda_b\langle b_s^{\dag}b_s\rangle+(\eta e^{-i\theta_a}\sinh r_a\cosh r_b-\eta^*e^{-i\theta_b}\sinh r_b\cosh r_a)\langle 
	a_s^{\dag}b_s^{\dag}\rangle-(\eta e^{i\theta_b}\sinh r_b\cosh r_a-\eta^*e^{i\theta_a}\notag\\
	&~~~~\times\sinh r_a\cosh r_b)\langle a_sb_s\rangle-(\eta^*\cosh r_a\cosh r_b-\eta e^{-i\Delta\theta}\sinh r_a\sinh r_b)\langle a_s^{\dag}b_s\rangle+(\eta^*e^{i\Delta\theta}\sinh r_a\sinh r_b\notag\\
	&~~~~-\eta \cosh r_a\cosh r_b)\langle a_sb_s^{\dag}\rangle+\Gamma |p_b|^2\sinh^2r_b\cosh^2r_c+\Gamma |p_b|^2\cosh^2r_b\sinh^2r_c+\Gamma [p_b^2e^{-i(\theta_c+\theta_b)}+p_b^{*2}\notag\\
	&~~~~\times e^{i(\theta_c+\theta_b)}]\sinh r_b\cosh r_b\sinh r_c\cosh r_c,\label{oe}
\end{align}
where $\Lambda_j=\Gamma|p_j|^2+\kappa_j$, $\Delta\theta=\theta_a-\theta_b$, $\zeta=iJe^{i\varphi}+\mu \frac{\Gamma}{2}$, $\eta=iJe^{-i\varphi}+\mu^* \frac{\Gamma}{2}$, and $\mu=p_a^*p_b$.

\section{Considerations on the squeezed-cavity frequency for battery energy} 
\begin{figure*}
	\centering
	\includegraphics[width=12cm,height=10.5cm]{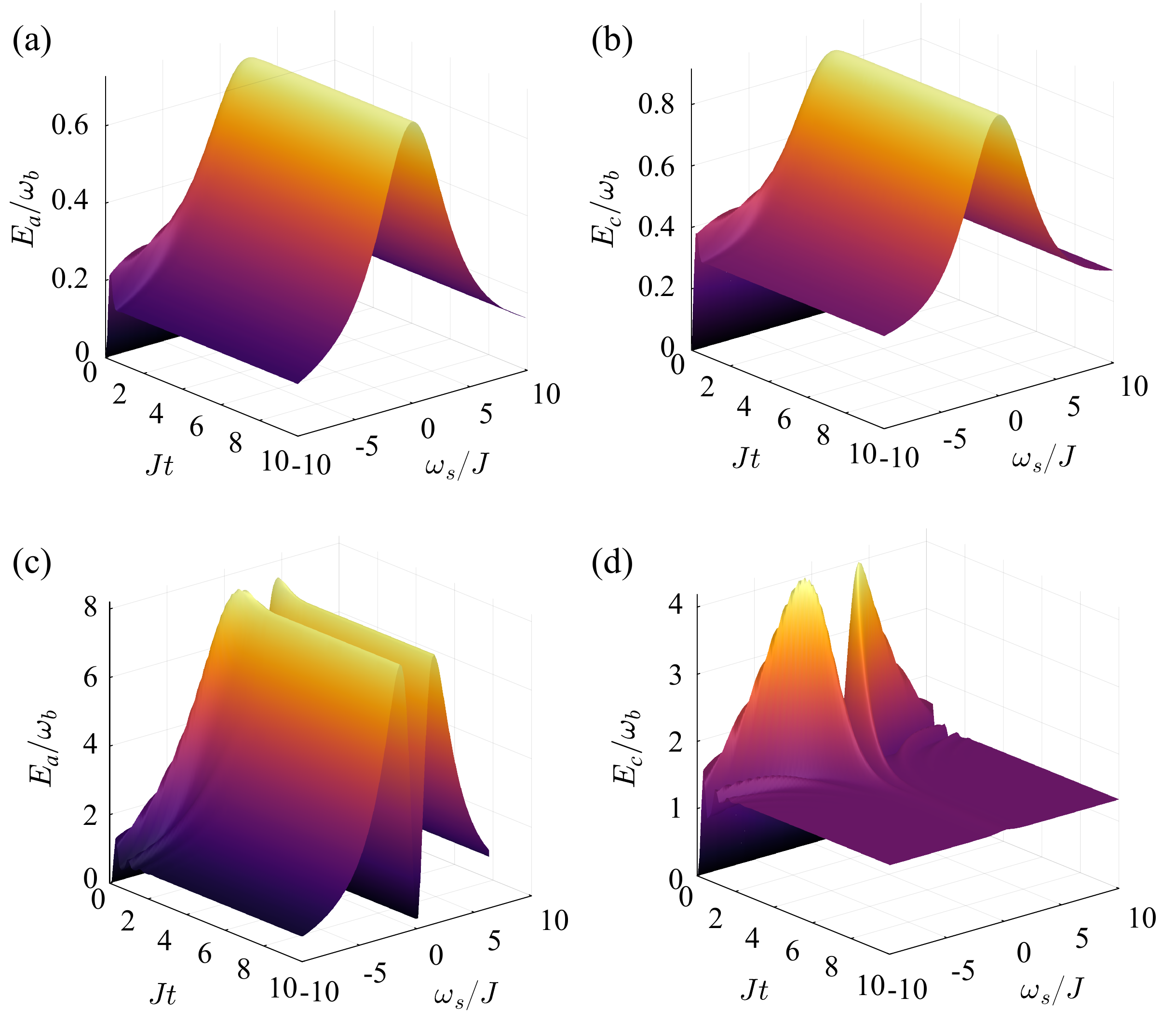}
	\caption{ Energy of quantum battery versus the scaled time $Jt$ and squeezed-cavity frequency $\omega_s$. (a,b) correspond to $J/\omega_b=1\times10^{-5}$ for case ($a$) and case ($c$), respectively, while (c,d) correspond to $J/\omega_b=0.001$. The parameters used are $\kappa=8\times 10^{-5}\omega_b$, $\epsilon= 10^{-4}\omega_b$, $r=1$, and $\Gamma=2J$.}
	\label{figs1}
\end{figure*}

For the quantum battery, a resonant classical drive field is applied to the charger, supplying energy to the system, which is described by the Hamiltonian $H_c=\epsilon(a+a^{\dag})$, where $\epsilon$ denotes the driving strength. After the squeezing transformation, the nonlinear Hamiltonian of the charger reduces to $H^a_{\rm NL}=\omega_sa_s^{\dag}a_s$, where $\omega_s$ denotes the squeezed-cavity frequency, i.e., there exists a detuning between the cavity modes $a_s$ and $b$. For cases ($a$) and ($c$),  we show the energy as a function of $\omega_s$ in Fig.~\ref{figs1}. For the Jaynes–Cummings model in the squeezing framework, the efficiency of the storage energy is determined by the detuning $\omega_s$ and coupling strength $J$. In the weak coupling regime, detuning suppresses the energy exchange between the two cavities, and the energy stored under resonance conditions is optimal, as shown in Figs.~\ref{figs1}(a, b). 

In contrast, in the strong coupling regime, due to the dissipation of the cavity, the resonance maximizes the exchange rate, but simultaneously enhances energy leakage through dissipative channels, thereby reducing the stored energy. In this case, a finite detuning can establish a balance between the exchange rate and dissipation, leading to the existence of an optimal detuning, as shown in Figs.~\ref{figs1}(c, d). For analytical convenience of the derivations and calculations, we set $\omega_s=0$ hereafter and in the main text. 

\section{Energy of the squeezing-improved nonreciprocal quantum battery} 
In this section, we first solve the equations of motion above to obtain analytical expressions for the battery energy under different cases, and then provide a detailed analysis of the underlying energy storage mechanism and the applicability of the model. For the case ($a$), the stored energy is given by
\begin{align}
	E_a&=\omega_b\langle b^{\dag}b\rangle =-\frac{4\omega_b J^2e^{-2r-(2\Lambda_a+\frac{3\Lambda_b}{2})t}}{\Lambda_a^2\Lambda_b^2\Lambda_-^2\Lambda_+}\Big\{(1+e^{4r})(2J-\Lambda_a)\big[\Lambda_a\Lambda_b\Lambda_-^2e^{(2\Lambda_a+\frac{3\Lambda_b}{2})t}+4\Lambda_a^2\Lambda_b^2e^{(\Lambda_b+\frac{3\Lambda_a}{2})t}\big]+32\epsilon^2\notag\\
	&~~~~\times\Lambda_+\Lambda_-e^{2r}\big[\Lambda_ae^{(2\Lambda_a+\Lambda_b)t}-\Lambda_be^{\frac{3}{2}\Lambda_+t}\big]+\Lambda_a\Lambda_b\Lambda_+(1+e^{4r})(\Lambda_a-2J)\big[\Lambda_ae^{(2\Lambda_a+\frac{1}{2}\Lambda_b)t}+\Lambda_be^{(\Lambda_a+\frac{3}{2}\Lambda_b)t}\big]\notag-2\\
	&~~~~\times e^{2r+(\Lambda_a+\frac{3\Lambda_b}{2})t}\Lambda_b^2\Lambda_+(8\epsilon^2-2J\Lambda_a+\Lambda_a^2)-2e^{2r+(2\Lambda_a+\frac{\Lambda_b}{2})t}\Lambda_a^2\Lambda_+(8\epsilon^2-2J\Lambda_b+\Lambda_a\Lambda_b)+8e^{2r+(\Lambda_b+\frac{3\Lambda_a}{2})t}\notag\\
	&~~~~\times\Lambda_a\Lambda_b(4\epsilon^2\Lambda_+-2J\Lambda_a\Lambda_b+\Lambda_a^2\Lambda_b)-2\Lambda_-^2e^{2r+(2\Lambda_a+\frac{3\Lambda_b}{2})t}(8\epsilon^2\Lambda_++2J\Lambda_a\Lambda_b-\Lambda_a^2\Lambda_b)\Big\},
\end{align}
where $\Lambda_{\pm}=\Lambda_a\pm\Lambda_b$, we have set $\omega_s=0$ and $r_a=r_c=r$ for simplicity.

For case ($b$), the stored energy is
\begin{align}
	E_b&=\frac{J\omega_be^{-(\frac{3\Lambda_a}{2}+\Lambda_b)t}}{\Lambda_a^2\Lambda_b^2\Lambda_-^2\Lambda_+}\Big\{128J\epsilon^2\Lambda_+\Lambda_-\big[\Lambda_be^{\Lambda_+t}-\Lambda_ae^{\frac{1}{2}(3\Lambda_a+\Lambda_b)t}\big]+16J\Lambda_b^2\Lambda_+e^{\frac{1}{2}t(\Lambda_a+2\Lambda_b)}(4\epsilon^2+J\Lambda_a)+\Lambda_a^2\Lambda_+e^{\frac{3t\Lambda_a}{2}}\notag\\
	&~~~~\times\big[64J\epsilon^2+(4J-\Lambda_a)^2\Lambda_b+2(4J-\Lambda_a)\Lambda_b^2+\Lambda_b^3\big]+16e^{\frac{1}{2}(2\Lambda_a+\Lambda_b)t}J\Lambda_a\Lambda_b\big[\Lambda_a\Lambda_b(\Lambda_--4J)-8\epsilon^2\Lambda_+\big]-\Lambda_-^2\notag\\
	&~~~~\times e^{(\frac{3}{2}\Lambda_a+\Lambda_b)t}(16J^2\Lambda_a\Lambda_b+\Lambda_a^2\Lambda_b\Lambda_+-8J\Lambda_a^2\Lambda_b-64J\epsilon^2\Lambda_+)-\Lambda_a\Lambda_b\big[16Je^{\frac{1}{2}(2\Lambda_a+\Lambda_b)t}\Lambda_a\Lambda_b(\Lambda_--4J)+16J^2\notag\\
	&~~~~\times\Lambda_b\Lambda_+e^{\frac{1}{2}(\Lambda_a+2\Lambda_b)t}+e^{\frac{3}{2}\Lambda_at}\Lambda_a\Lambda_+(4J-\Lambda_-)^2\big]-\Lambda_a\Lambda_b\Lambda_-^2e^{(\frac{3}{2}\Lambda_a+\Lambda_b)t}\big[(-4J+\Lambda_a)^2+\Lambda_a\Lambda_b\big]\cosh{2r}\Big\}.
\end{align}

And for case ($c$), the energy is
\begin{align}
	E_c&=\frac{e^{-\Lambda_bt}}{2\Lambda_a^2\Lambda_b^2\Lambda_-^2\Lambda_+}\Big\{\Lambda_+\big[(1-e^{\Lambda_bt})\Gamma\Lambda_a^2\Lambda_-^2\Lambda_b+16J^2(e^{-\frac{1}{2}\Lambda_-t}-1)^2\Lambda_a^2\Lambda_b^2+128J^2e^{-\Lambda_at}\epsilon^2(\Lambda_ae^{\frac{1}{2}\Lambda_at}-\Lambda_be^{\frac{1}{2}\Lambda_bt}\notag\\
	&~~~~-\Lambda_-\Lambda_ae^{\frac{1}{2}\Lambda_+t})^2\big]+\Lambda_a\Lambda_b\big[(e^{\Lambda_bt}-1)\Gamma\Lambda_a\Lambda_-^2\Lambda_++8J^3(\Lambda_a\Lambda_+-\Lambda_-^2e^{\Lambda_bt}-4\Lambda_a\Lambda_be^{-\frac{1}{2}\Lambda_-t}+\Lambda_b\Lambda_+e^{-\Lambda_-t})\notag\\
	&~~~~-16J^2(e^{-\frac{1}{2}\Lambda_-t}-1)^2\Lambda_a\Lambda_b\Lambda_++4J^2\Gamma(\Lambda_-^2e^{\Lambda_bt}+4\Lambda_a\Lambda_be^{-\frac{1}{2}\Lambda_-t}-\Lambda_a^2-\Lambda_b\Lambda_+e^{-\Lambda_-t})\big]\cosh{2r}+4J^2\Lambda_a\notag\\
	&~~~~\times\Lambda_b(2J-\Gamma)(\Lambda_-^2e^{\Lambda_bt}+4\Lambda_a\Lambda_be^{-\frac{1}{2}\Lambda_-t}-\Lambda_a\Lambda_+-\Lambda_b\Lambda_+e^{-\Lambda_-t})\cosh{6r}\Big\}.
\end{align}
\subsection{Analysis of energy storage mechanisms}
Dissipation drives the system to a steady state; the corresponding stored energy is given by Eq.~(5) in the main text. Here, we present a detailed analysis of how squeezing the charger and the reservoir affects the battery energy. For the case of only squeezing the charger, the master equation of the system can be expanded as
\begin{align}
	\dot\rho_s(t)=&-i[H_s,\rho_s(t)]+\sum_{j=a,b}S_a^{\dag}\mathcal{L}[L_j]\rho(t)S_a+S_a^{\dag}\mathcal{L}[L_c]\rho(t)S_a\notag\\
	=&-i[H_s,\rho_s(t)]+\sum_{j=a_s,b}\mathcal{L}[L_j]\rho_s(t)+\Gamma\big[a_s\rho_s(t)a^{\dag}_s-1/2\{a^{\dag}_sa_s,\rho_s(t)\}\big]+\Gamma\big[b\rho_s(t)b^{\dag}-1/2\{b^{\dag}b,\rho_s(t)\}\big]\notag\\
	&+\Gamma p_ap_b^*\mathcal{D}(b,a_s)\rho_s(t)+\Gamma p_ap_b^*\mathcal{D}(a_s,b)\rho_s(t)
	\notag\\
	=&-i[H_s,\rho_s(t)]+\sum_{j=a_s,b}\mathcal{L}[L'_j]\rho_s(t)+\Gamma p_ap_b^*\mathcal{D}(b,a_s)\rho_s(t)+\Gamma p_a^*p_b\mathcal{D}(a_s,b)\rho_s(t),
\end{align}
where $\mathcal{D}(A,B)$ describes the cooperative dissipative coupling between the charger and battery, whose interplay with the coherent coupling gives rise to nonreciprocity. In addition, $L'_{a_s(b)}=\sqrt{\kappa_{a(b)}+\Gamma}a_s(b)$ indicates that the common reservoir also enhances the local dissipation rates. There exists an optimal $J_{\mathrm{op}}$ that maximizes the steady-state energy.

We note that squeezing the reservoir gives rise to a qualitatively different behavior. Here we provide a detailed analysis where only the common reservoir is squeezed. The pair correlations in the squeezed vacuum reservoir increase exponentially with the squeezing parameter, allowing the system to exchange excitations in correlated pairs, which does not modify the \emph{first}-order dynamics, but instead reshapes the \emph{higher}-order dynamics of the system.

Indeed, from the equations of motion derived in Eq.~(\ref{oe}), setting 
$r_a=r_b=0$ yields the following expressions for the \emph{second}-order moments:
\begin{align}
	\frac{d\langle a^{\dag}a\rangle}{dt}&=-\Lambda_a\langle a^{\dag}a\rangle-\zeta\langle a^{\dag}b\rangle-\zeta^*\langle ab^{\dag}\rangle+i\epsilon\langle a\rangle-i\epsilon\langle a^{\dag}\rangle+\Gamma |p_a|^2\sinh^2r_c,\notag\\
	\frac{d\langle a^{\dag}b\rangle}{dt}&=-\frac{\Lambda_a+\Lambda_b}{2}\langle a^{\dag}b\rangle-\eta\langle a^{\dag}a\rangle-\zeta^*\langle b^{\dag}b\rangle+i\epsilon\langle b\rangle+\mu^*\Gamma \sinh^2r_c,\notag\\
	\frac{d\langle b^{\dag}b\rangle}{dt}&=-\Lambda_b\langle b^{\dag}b\rangle-\eta^*\langle a^{\dag}b\rangle-\eta\langle ab^{\dag}\rangle+\Gamma |p_b|^2\sinh^2r_c.\label{S18}
\end{align}
Squeezing the common reservoir introduces a constant gain term $(\Gamma |p_b|^2\sinh^2r_c)$ in the equation for $\langle b^{\dag}b\rangle$, which originates from the effective occupied number $N=\sinh^2r_c$ of the squeezed reservoir. Since this constant contribution exists independently of the buildup of coherence and correlations, it enhances the effective energy transfer rate into the battery, thereby significantly increasing the stored energy and enabling faster charging. But for the squeezing of the charger, the gain term affects the battery energy only indirectly through the coherent coupling between operators; the battery acquires energy only after the relevant coherence
and correlations are established. Moreover, the squeezed reservoir not only increases system energy but also provides ordered quantum coherent resources. In particular, the anomalous correlation $M$ introduces phase-sensitive quantum coherence that cannot be generated by a thermal reservoir. These coherence modify the system dynamics and enhance the extractable work within the system.

\begin{figure*}
	\centering
	\includegraphics[width=17.2cm,height=4.6cm]{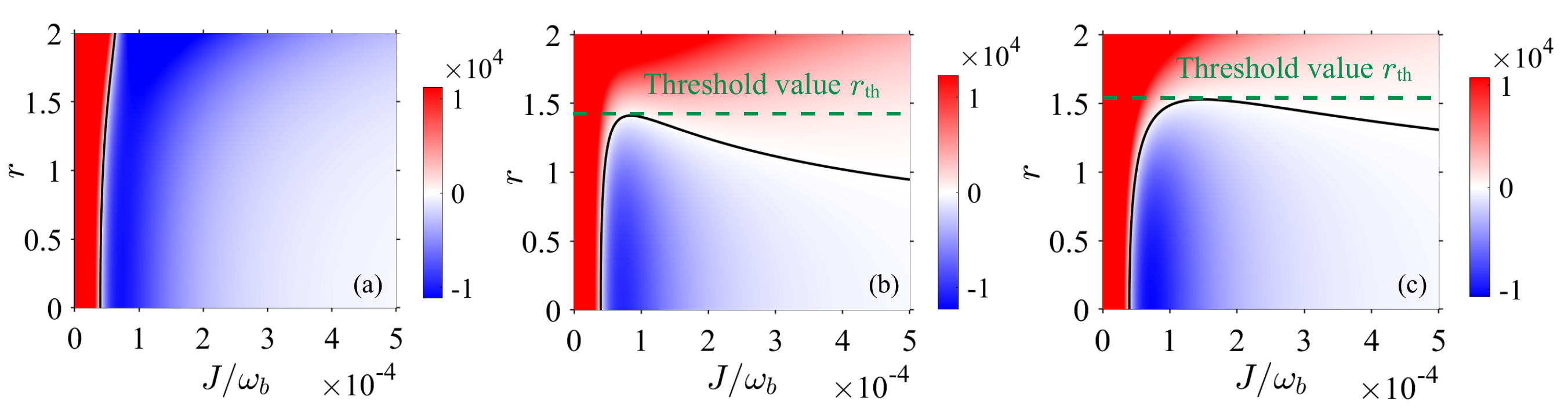}
	\caption{Partial derivative $\partial E_i^{\rm ss}/\partial J$ of the steady-state energy $E_i^{\rm ss}$ versus the coherent coupling strength $J$ and the squeezing parameter $r$ for case ($a$), case ($b$), and case ($c$).}
	\label{figss}
\end{figure*}

For cases ($b$) and ($c$), a threshold squeezing strength $r_{\rm{th}}$ emerges, above which the steady-state energy increases monotonically with the coupling strength $J$. This behavior originates from the fact that the dissipative channels inject energy into the system. As the squeezing strength increases, these contributions become dominant, masking the competition between coherent coupling and dissipative coupling. For all cases ($a,b,c$), the derivative of the energy takes the form
\begin{align}
	\frac{\partial E_a^{\rm ss}}{\partial J}&=\frac{16J[8\epsilon^2(-2J+\kappa)+\kappa(-J+\kappa)(2J+\kappa)\sinh^2r]}{(2J+\kappa)^5},\notag\\
	\frac{\partial E_b^{\rm ss}}{\partial J}&=\frac{128J\epsilon^2(-2J+\kappa)+2\kappa(2J+\kappa)(12J^2-4J\kappa+\kappa^2)\sinh^2r}{(2J+\kappa)^5},\notag\\
	\frac{\partial E_c^{\rm ss}}{\partial J}&=\frac{2[64J\epsilon^2(-2J+\kappa)+\kappa(2J+\kappa)^3\sinh^2r]}{(2J+\kappa)^5}.
\end{align}
As shown in Fig.~\ref{figss}, it remains zero over the entire parameter range when $r\textgreater r_{\rm{th}}$, indicating a strictly monotonic dependence of the steady-state energy on $J$.

\begin{figure*}
	\centering
	\includegraphics[width=15cm,height=4.04cm]{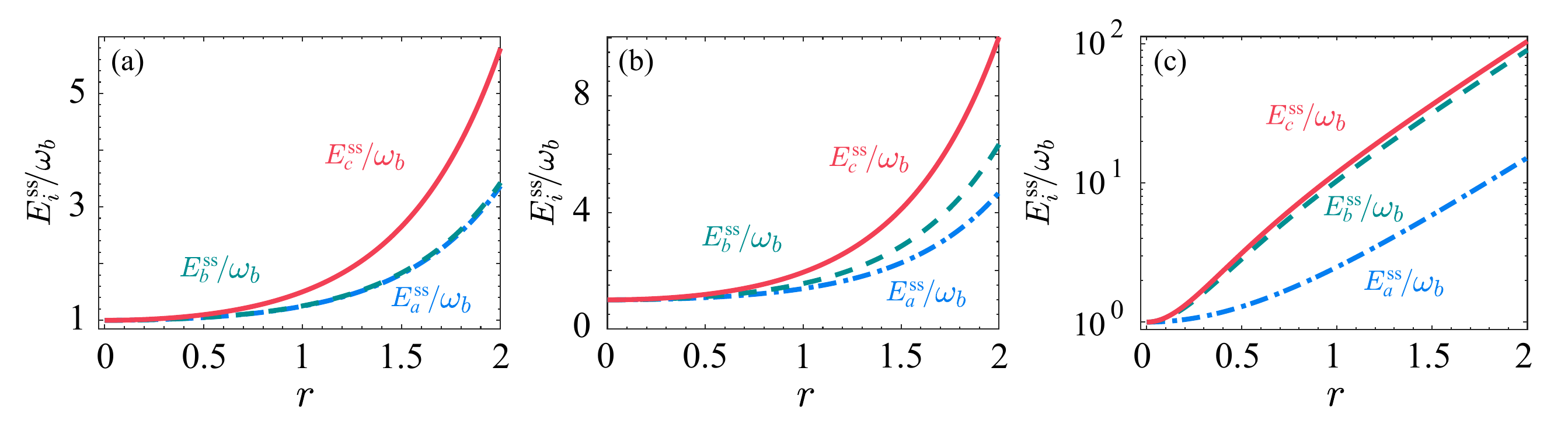}
	\caption{Steady-state energy versus the squeezing parameter $r$ for different coupling strengths, with $J/\omega_b=5\times 10^{-5}$ for (a), $J/\omega_b=10^{-4}$ for (b), and $J/\omega_b=5\times 10^{-4}$ for (c).}
	\label{figs}
\end{figure*}

Furthermore, the enhancement of nonreciprocal coupling represents only one aspect of the squeezing-induced effects. In the squeezed framework, additional counter-rotating contributions naturally arise and and the classical driving field is amplified to $\epsilon[(\cosh{r_a}-e^{i\theta_a}\sinh{r_a})a_s+(\cosh{r_a}-e^{-i\theta_a}\sinh{r_a})a_s^{\dag}]$. Figure~\ref{figs} shows that, in both the strong- and weak-coupling regimes, the stored energy increases monotonically with the squeezing strength $r$, reflecting the cooperative enhancement of effective energy-injection channels induced by squeezing.

\subsection{Limitations of energy storage}
According to Eq. (5) in the main text, the steady-state energies can be expressed as
\begin{align}
	E_i^{\rm ss}&=A_i\sinh^2 r+E^{\rm ss},~~~~(i=a,b, \mathrm{or}~c),\notag
\end{align}
with
\[
A_a=\frac{8J^2\kappa}{(2J+\kappa)^3},\quad
A_b=\frac{2J(4J^2+\kappa^2)}{(2J+\kappa)^3},\quad
A_c=\frac{2J}{2J+\kappa}.
\]
For the coherent coupling strength $J$, the energy stored in the battery exhibits an optimum coupling strength $J_{\mathrm{op}}$, at which the stored energy reaches a maximum [see the white dashed lines in Fig.~3(a-c) of the main text]. This can be understood from the competition between coherent energy exchange and dissipation, leading to an effective matching condition. In contrast, for the squeezing parameter, the derivative of the stored energy with respect to $r$ is
\[
\frac{\partial E_i^{\rm ss}}{\partial r}=A_i\sinh(2r)>0,
\]
which is always positive. The correlation coefficients of the squeezing vacuum reservoir are related to $N$ and $M$, both of which increase monotonically and are unbounded with respect to the squeezing parameter. As a result, the squeezing-induced contributions in the master equation grow with $r$, and the corresponding stored energy does not exhibit saturation.

These conclusions are derived under the assumption of a broadband squeezed reservoir, where the reservoir correlation time is much shorter than all relevant system timescales. However, in realistic physics platforms, the stored energy may be limited by imperfections of the squeezing source and of the propagation/coupling channels, including finite pump power and pump depletion, optical or microwave losses, phase noise, excess thermal noise or heating, and the finite bandwidth of the engineered reservoir. In particular, loss and phase noise degrade the observable squeezing, while pump depletion causes the pump field to evolve dynamically, thereby modifying the reservoir correlations. Moreover, when the squeezing bandwidth is no longer sufficiently large compared with the relevant spectral scales of the system, reservoir temporal correlations cannot be neglected, and the broadband Markovian description no longer holds. In this regime, a more complicated model incorporating finite bandwidth and pump dynamics would be required.  Alternatively, these effects can be mitigated via appropriate engineering techniques, allowing the system to still be described well within a Markovian framework. Exploring these directions remains worthy for future exploration.

\section{Analytical calculation for the ergotropy of the quantum battery} 
Apart from the energy storage performance, the extractable energy~(ergotropy) is also an important criterion for evaluating batteries, defined as $\mathcal{E}_i^{\rm ss}=E_i^{\rm ss}-\widetilde{E}_i^{\rm ss}=\mathrm{Tr}[\rho_b(t)H_b]-\mathrm{Tr}[\widetilde{\rho}_b(t)H_b]$. Here, $E_i^{\rm ss}$ denotes the stored energy of the battery, $\widetilde{\rho}_b$ is the passive state, from which no work can be extracted via unitary cyclic processes. Explicitly, if $\rho_b=\sum_nr_n|r_n\rangle\langle r_n|$, with $r_1\geq r_2\geq\cdots$, and $H_b=\sum_n\epsilon_n|\epsilon_n\rangle\langle\epsilon_n|$ with $\epsilon_1\leq\epsilon_2\leq\cdots$, then $\widetilde{\rho}_b=\sum_nr_n|\epsilon_n\rangle\langle\epsilon_n|$. Our numerical results are obtained by solving the master equation and substituting the resulting density matrix into the above expression. For the solution of analytical expression, the corresponding passive energy $\widetilde{E}_i^{\rm ss}$ can be expressed as~\cite{PhysRevB.99.035421,Downing2023}
\begin{align}\label{s12}
	\widetilde{E}_i^{\rm ss}=\omega_b\big(\frac{\sqrt{\mathcal{J}_i}-1}{2}\big),
\end{align}
with
\begin{align}
	\mathcal{J}=\big(1+2\langle b^{\dag}b\rangle-2\langle b^{\dag}\rangle\langle b\rangle\big)^2-4\big|\langle bb\rangle-\langle b\rangle^2\big|^2.\notag
\end{align}
By solving the equation of motion for the operators, the ergotropy for the battery can also be obtained; the corresponding $\mathcal{J}$ is
\begin{align}
	\mathcal{J}_a=&\frac{256J^4\sinh^2r}{\Lambda^8}\Big\{\big[8(1-i)\epsilon^2+(2J-\Lambda)\Lambda\big]\cosh r-8(1-i)\epsilon^2\sinh r\Big\}\Big\{\big[-8(1+i)\epsilon^2-(2J-\Lambda)\Lambda\big]\cosh r\notag\\
	&+8(1+i)\epsilon^2\sinh r\Big\}+\frac{1}{\Lambda^6}\big[-8J^2\Lambda+\Lambda^3+8J^2(\Lambda \cosh{2r}-4J\sinh^2r)\big]^2,\notag\\
	\mathcal{J}_b=&\frac{1}{\Lambda^8}\Big\{\Lambda^2\big[(2J-\Lambda)(8J^2+\Lambda^2)-2J(8J^2-4J\Lambda+\Lambda^2)\cosh{2r}\big]^2-J^2\big[64J\epsilon^2(1-\cosh r)+(2\Lambda^3+16J^2\Lambda\notag\\
	&-8J\Lambda^2)\sinh{2r}\big]^2\Big\},\notag\\
	\mathcal{J}_c=&\frac{1}{\Lambda^6}\Big\{4J\Lambda^4(\Lambda-2J)\cosh{2r}-(8J^2-4J\Lambda+\Lambda^2)\big[-32J^4+16J^3\Lambda-\Lambda^4+16J^3(2J-\Lambda)\cosh{4r}\big]\Big\},
\end{align}
where we have set $\Lambda_a=\Lambda_b=\Lambda$ for simplicity. Substituting it into Eq.~(\ref{s12}), it is easy to obtain the analytical solution of $\mathcal{E}_i^{\rm ss}$ for different cases.

\begin{figure*}
	\centering
	\includegraphics[width=10cm,height=4.24cm]{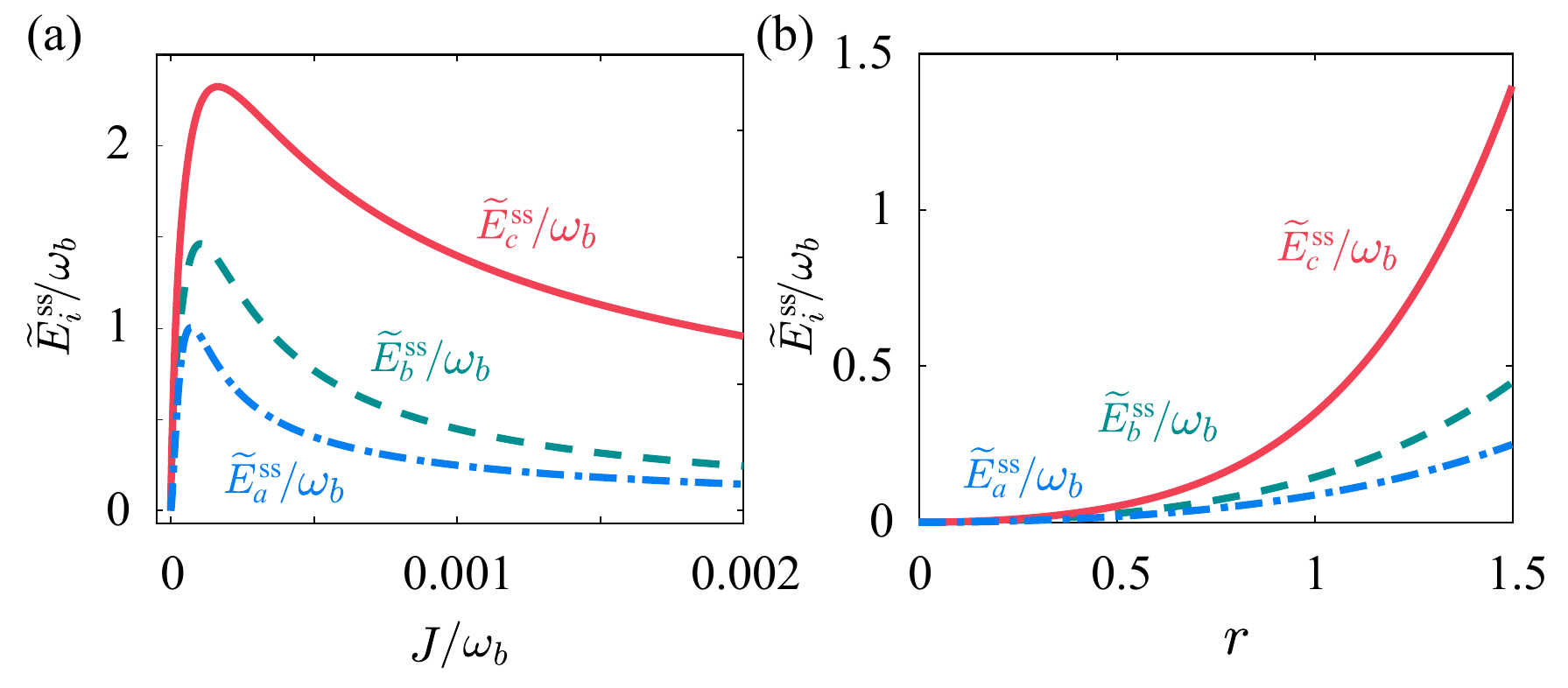}
	\caption{Energy $\widetilde{E}_i^{\rm ss}$ of the passive state as the function of the coupling strength $J$ (a) and squeezing parameter $r$ (b) in the steady state. The parameters used are $\kappa=8\times 10^{-5}\omega_b$, $\epsilon= 10^{-4}\omega_b$, $r=1.5$,  $J/\omega_b=0.001$, and $\Gamma=2J$.}
	\label{figs3}
\end{figure*}

The energy of the passive state is illustrated in Fig.~\ref{figs3}. At the optimal coupling strength $J$, the corresponding $\widetilde{E}_i^{\rm ss}$ also attains its maximal value. Moreover, $\widetilde{E}_i^{\rm ss}$ increases monotonically with the squeezing parameter $r$. This result indicates that the increase in steady-state energy does not necessarily lead to an increase in ergotropy. Squeezing of the reservoir can stably maintain coherence in the steady state, maximizing the ergotropy of the battery; while the additional battery squeezing increases energy, it mainly introduces a passive state, which has a limited contribution to coherence or even partially counteracts it, thus the ergotropy in case ($c$)  decreases and even becomes lower than case ($b$) with the increase of $r$.

It is worth noting that small deviations between the analytical and numerical results in the main text arise from the different approximations employed in the two approaches. In the numerical simulations, a finite-dimensional truncation of the Hilbert space is introduced, while the analytical treatment is obtained within a second-order mean-field approximation.
\section{Scattering coefficient matrix of the optical isolator}
When both cavity modes are squeezed, the corresponding quantum Langevin equations~(QLEs) can be found in Eq.~(6) in the main text. To analyze the transmission properties, we linearize the QLEs as $dV/dt=-MV+D V_{\mathrm{in}}$, where $V=(a_s,b_s,a_s^{\dag},b_s^{\dag})^T$ is the vector of system operators, $V_{\mathrm{in}}=(a^{\mathrm{in}}, b^{\mathrm{in}}, a^{\mathrm{in}\dag}, b^{\mathrm{in}\dag})^T$, and $D=\mathrm{diag}(\sqrt{\kappa_a}, \sqrt{\kappa_b}, \sqrt{\kappa_a}, \sqrt{\kappa_b})$. The coefficient matrix $M$ can be written as
\begin{align}
	M=\begin{pmatrix}
		-\frac{\Lambda_a+i\omega_s}{2}&0&0&0\\
		2J(e^{i\Delta\theta}\sinh r_a\sinh r_b-\cosh r_a\cosh r_b)&-\frac{\Lambda_b+i\omega_s}{2}&2J(e^{-i\theta_a}\sinh r_a\cosh r_b-e^{-i\theta_b}\cosh r_a\sinh r_b)&0\\
		0&0&-\frac{\Lambda_a-i\omega_s}{2}&0\\
		2J(e^{i\theta_a}\sinh r_a\cosh r_b-e^{i\theta_b}\cosh r_a\sinh r_b)&0&2J(e^{-i\Delta\theta}\sinh r_a\sinh r_b-\cosh r_a\cosh r_b)&\frac{i\omega_s-\Lambda_b}{2}
	\end{pmatrix}.\notag
\end{align}
Transforming to the frequency domain yields
\begin{align}
	\overline{V}(\omega)=(M-i\omega \mathbb{I})^{-1}D \overline{V}_{\mathrm{in}}(\omega).
\end{align}
Using the input-output relation $\nu_{\mathrm{out}}+\nu_{\mathrm{in}}=\sqrt{\gamma_{\nu}}\nu$~\cite{PhysRevA.31.3761}, the output field is $\overline{V}_{\mathrm{out}}(\omega)=U(\omega)\overline{V}_{\mathrm{in}}(\omega)$, where 
\begin{align}
	\overline{V}_{\mathrm{out}}=\big(a^{\mathrm{out}}(\omega), b^{\mathrm{out}}(\omega), a^{\mathrm{out}\dag}(\omega), b^{\mathrm{out}\dag}(\omega)\big)^T, 
\end{align}
and
\begin{align}
	U(\omega)=D(M-i\omega \mathbb{I})^{-1}D-\mathbb{I},  
\end{align} 
which can be expressed as
\begin{align}
	\begin{pmatrix}
		-\frac{2\kappa_a}{\Lambda_a+i(2\omega+\omega_s)}&0&0&0\\
		\frac{8J\sqrt{\kappa_a\kappa_b}(\cosh r_a\cosh r_b-e^{i\Delta\theta}\sinh r_a\sinh r_b)}{[\Lambda_a+i(2\omega+\omega_s)][\Lambda_b+i(2\omega+\omega_s)]}&\frac{-2\kappa_b}{\Lambda_b+i(2\omega+\omega_s)}&\frac{8J\sqrt{\kappa_a\kappa_b}(e^{-i\theta_b}\cosh r_a\sinh r_b-e^{-i\theta_a}\sinh r_a\cosh r_b)}{(i\Lambda_a-2\omega+\omega_s)(-i\Lambda_b+2\omega+\omega_s)}&0\\
		0&0&-\frac{2\kappa_a}{\Lambda_a+i(2\omega-\omega_s)}&0\\
		\frac{8J\sqrt{\kappa_a\kappa_b}(e^{i\theta_b}\cosh r_a\sinh r_b-e^{i\theta_a}\sinh r_a\cosh r_b)}{(i\Lambda_b-2\omega+\omega_s)(-i\Lambda_a+2\omega+\omega_s)}&0&\frac{8J\sqrt{\kappa_a\kappa_b}(\cosh r_a\cosh r_b-e^{-i\Delta\theta}\sinh r_a\sinh r_b)}{[\Lambda_a+i(2\omega-\omega_s)][\Lambda_b+i(2\omega-\omega_s)]}&\frac{-2\kappa_b}{\Lambda_b+i(2\omega-\omega_s)}
	\end{pmatrix}-\mathbb{I}.\label{eq26}
\end{align}

Substituting 
\begin{align}
	a_s^{\mathrm{out}}(\omega)=U_{11}(\omega)a_s^{\mathrm{in}}(\omega)+U_{12}(\omega)b^{\mathrm{in}}(\omega)+U_{13}(\omega)a_s^{\mathrm{in}\dag}(\omega)+U_{14}(\omega)b^{\mathrm{in}\dag}(\omega),\notag\\
	b^{\mathrm{out}}(\omega)=U_{21}(\omega)a_s^{\mathrm{in}}(\omega)+U_{22}(\omega)b^{\mathrm{in}}(\omega)+U_{23}(\omega)a_s^{\mathrm{in}\dag}(\omega)+U_{24}(\omega)b^{\mathrm{in}\dag}(\omega),
\end{align}
and their Hermitian conjugate into the definition of the output field spectrum, $s_{\nu}^{\mathrm{out}}(\omega)=\int d\omega'\langle\overline{\nu}^{\mathrm{out}\dag}(\omega')\overline{\nu}^{\mathrm{out}}(\omega)\rangle$~\cite{PhysRevA.85.021801}, we can obtain 
\begin{align}
	s_{a}^{\mathrm{out}}(\omega)=\int d \omega'\sum_{j,k=1}^4U_{1j}^*(\omega')U_{1k}(\omega)\langle\overline{V}_{\mathrm{in},j}^{\dag}(\omega')\overline{V}_{\mathrm{in},k}(\omega)\rangle,\notag\\
	s_{b}^{\mathrm{out}}(\omega)=\int d \omega'\sum_{j,k=1}^4U_{2j}^*(\omega')U_{2k}(\omega)\langle\overline{V}_{\mathrm{in},j}^{\dag}(\omega')\overline{V}_{\mathrm{in},k}(\omega)\rangle.
\end{align}
The subscript of $U(\omega)$ indicates its matrix element. Utilizing the correlation of the input field $\langle \nu^{\mathrm{in}\dag}(\omega')\nu^{\mathrm{in}}(\omega)\rangle=s_{\nu}^{\mathrm{in}}(\omega)\delta(\omega-\omega')$, we can obtain 
\begin{align}
	s_{a}^{\mathrm{out}}(\omega)=\big(|U_{11}|^2+|U_{13}|^2\big)s_{a}^{\mathrm{in}}(\omega)+\big(|U_{12}|^2+|U_{14}|^2\big)s_{b}^{\mathrm{in}}(\omega)+|U_{13}|^2+|U_{14}|^2,\notag\\
	s_{b}^{\mathrm{out}}(\omega)=\big(|U_{21}|^2+|U_{23}|^2\big)s_{a}^{\mathrm{in}}(\omega)+\big(|U_{22}|^2+|U_{24}|^2\big)s_{b}^{\mathrm{in}}(\omega)+|U_{23}|^2+|U_{24}|^2.\end{align}

Furthermore, the spectrum of the output field can be expressed as \begin{align}
	S_{\mathrm{out}}(\omega)=T(\omega)S_{\mathrm{in}}(\omega)+S_{\mathrm{vac}}(\omega), 
\end{align}
with $S_{\mathrm{in}}(\omega)=\big(s_{a}^{\mathrm{in}}(\omega), s_{b}^{\mathrm{in}}(\omega)\big)^T$, $S_{\mathrm{out}}(\omega)=\big(s_{a}^{\mathrm{out}}(\omega), s_{b}^{\mathrm{out}}(\omega)\big)^T$, $S_{\mathrm{vac}}(\omega)=\big(s_{a}^{\mathrm{vac}}(\omega), s_{b}^{\mathrm{vac}}(\omega)\big)^T$, and
\begin{align}
	T(\omega)=
	\begin{pmatrix}
		|U_{11}|^2+|U_{13}|^2&|U_{12}|^2+|U_{14}|^2\\
		|U_{21}|^2+|U_{23}|^2&|U_{22}|^2+|U_{24}|^2
	\end{pmatrix},
\end{align}
$s_{\nu}^{\mathrm{vac}}$ is the output spectrum contributing from the
input vacuum field, which is given by $s_{a}^{\mathrm{vac}}=|U_{13}|^2+|U_{14}|^2$ and $s_{b}^{\mathrm{vac}}=|U_{23}|^2+|U_{24}|^2$. With Eq.~(\ref{eq26}), the transmission coefficient of the signal from mode $b$ to mode $a$ vanishes, $T_{ab}=0$, whereas the transmission coefficient from mode $a$ to mode $b$ is
\begin{align}
	T_{ba}
	=&\frac{64J^2\kappa_a\kappa_b}{\Lambda_b^2+(2\omega+\omega_s)^2}
	\left[
	\frac{
		2+\cosh\!\bigl[2(r_a-r_b)\bigr]
		+\cosh\!\bigl[2(r_a+r_b)\bigr]
		-2\cos\Delta\theta\sinh2r_a\sinh2r_b
	}{
		4\bigl[\Lambda_a^2+(2\omega+\omega_s)^2\bigr]
	}
	\right. \notag\\
	&\left.
	+\frac{
		\sinh^2 r_a \cosh^2 r_b + \cosh^2 r_a \sinh^2 r_b
		-\tfrac{1}{2}\cos\Delta\theta\sinh2r_a\sinh2r_b
	}{
		\Lambda_a^2+(-2\omega+\omega_s)^2
	}
	\right].
\end{align}
	
\end{document}